\documentclass[final,3p]{elsarticle}
\biboptions{numbers,sort&compress}

\usepackage{amssymb}
\usepackage{hyperref}
\usepackage[load=addn,load=physical]{siunitx}
\usepackage{multirow}
\usepackage{wasysym}
\usepackage{xcolor}
\usepackage{subfigure}
\usepackage[utf8]{inputenc}

\journal{Nucl. Instrum. Meth. A}

\begin{document}

\begin{frontmatter}



\title{Combined analysis of HPK 3.1 LGADs using a proton beam, beta source, and probe station towards establishing high volume quality control}


\author[1]{R.~Heller\corref{cor}}\ead{rheller@fnal.gov}

\author[3]{A.~Abreu}
\author[1]{A.~Apresyan}
\author[6,8]{R.~Arcidiacono}
\author[6]{N.~Cartiglia}
\author[1]{K.~DiPetrillo}
\author[6]{M.~Ferrero}
\author[1]{M.~Hussain} 
\author[3]{M.~Lazarovitz}
\author[4]{H.~Lee}
\author[1]{S.~Los}
\author[4]{C.S.~Moon}
\author[1,2]{C.~Peña}
\author[7]{F.~Siviero}
\author[6]{V.~Sola}
\author[5]{T.~Wamorkar}
\author[2]{S.~Xie}

\address[1]{Fermi National Accelerator Laboratory, Batavia, IL, USA}
\address[2]{California Institute of Technology, Pasadena, CA, USA}
\address[3]{University of Kansas, KS, USA}
\address[4]{Kyungpook National University, Daegu, South Korea}
\address[5]{Northeastern University, MA, USA}
\address[6]{INFN, Torino, Italy}
\address[7]{Universit\`a di Torino, Torino, Italy}
\address[8]{Universit\`a del Piemonte Orientale, Italy}
\cortext[cor]{Corresponding author}

\begin{abstract}
The upgrades of the CMS and ATLAS experiments for the high luminosity phase of the Large Hadron Collider will employ precision timing detectors based on Low Gain Avalanche Detectors (LGADs). We present a suite of results combining measurements from the Fermilab Test Beam Facility, a beta source telescope, and a probe station,
allowing full characterization of the HPK type 3.1 production of LGAD prototypes developed for these detectors. We demonstrate that the LGAD response to high energy test beam particles is accurately reproduced with a beta source.
We further establish that probe station measurements of the gain implant accurately predict the particle response and operating parameters of each sensor, and conclude that the uniformity of the gain implant in this production is sufficient to produce full-sized sensors for the ATLAS and CMS timing detectors.
\end{abstract}


\end{frontmatter}

\tableofcontents




\section{Introduction} 

Future colliders, including the high luminosity upgrade of the Large Hadron 
Collider (HL-LHC) at CERN, will operate with instantaneous luminosities at 
least five times higher than current LHC running conditions. The rise in 
instantaneous luminosity will increase the rate of simultaneous interactions 
per bunch crossing (pileup) to approximately 200 at the 
HL-LHC~\cite{Apollinari:2120673}, and up to 1,200 at the FCC-hh~\cite{Schulte:2017qkc}. 
The large amount of pileup exacerbates the difficulties in separating 
particles that originate from the hard-scattering from those produced 
in pileup interactions. Precision timing has been identified as a 
pileup mitigation technique to complement precision tracking at the 
HL-LHC. Pileup interactions are spread over a period of 
approximately \SI{200}{\pico \second}, and a time measurement with precision 
of approximately 30-40~\si{\pico \second} will reduce the effective rate of 
pileup by a factor of five, yielding pileup levels comparable with current LHC 
conditions. In this paper, we report results of measurements with thin 
low-gain avalanche detectors (LGADs), which will be used in the CMS and 
ATLAS upgrades for HL-LHC~\cite{CMS:2667167} and have been demonstrated 
to achieve time resolutions below 
30~\si{ps}~\cite{Cartiglia201783, PELLEGRINI201412,ApresyanLGAD}.

Previous measurements of LGAD sensor performance have focused on 
small, \si{\milli\m} scale devices with few pads. The CMS timing detector 
will be constructed of larger area 
sensors (\SI[product-units = power]{2 x 4}{\centi \meter}). 
Effective operation of such large sensors will require highly uniform gain layer 
deposition to enable the required response across the sensor area constrained 
to a single operating voltage. The ability to produce uniform LGAD wafers 
is thus a key requirement for a successful timing detector. 
The measurements presented in this paper are among the first to explicitly
address this scalability requirement.

In this paper, we follow several approaches to study the LGAD batch produced in 2018 by Hamamatsu, hereafter referred to as HPK type 3.1 LGADs. Enabled by the development 
of a 16-channel readout board at Fermilab, we demonstrate for the first time 
the successful operation of a large, 16-pad LGAD sensor 
using \SI{120}{\GeV} protons at the Fermilab Test Beam Facility (FTBF). These
measurements demonstrate an HPK type 3.1 LGAD sensor that meets the uniformity in efficiency, gain, and time resolution that are required for the CMS and ATLAS timing detectors. Furthermore, we demonstrate that the 
size of the inactive interpad-gap regions measured by a laser using the
transient current technique (TCT) is consistent with the proton beam measurement.

We then establish that sensor characterization based on a Ruthenium-106 beta source yields results consistent with those from the proton beam. This conclusion enables the use of the beta source setup to perform studies of sensor uniformity on scales across the entire wafer. These high-volume studies would not be feasible with the proton beam alone, due to the limited availability of beamtime. Our measurements demonstrate that the gain uniformity across the 
HPK type 3.1 LGAD wafer meets the requirements of the CMS and ATLAS timing detectors.

Finally, by comparing beta source measurements with probe station measurements 
of the gain implant in each sensor, we establish a clear linear relationship
between the operational bias voltage and the gain layer depletion voltage. This observation allows us to reliably predict the gain
of the sensor from probe station capacitance-voltage (CV) measurements,
critically enhancing the power of the probe station as a tool for studying 
wafer-scale uniformity. As a result, we show that the probe station 
can be used as a fast and reliable way to establish sensor quality 
control during the production phase of the CMS and ATLAS timing detectors.  
A similar correspondence between the gain layer depletion voltage and the operating voltage has been established for wide variation between distinct sensor designs~\cite{JIN2020164611}. In contrast, the results shown here probe this relationship with a larger sensor population, and focus on variation within the manufacturing tolerance for the context of production quality control. 

The paper is organized as follows: the LGAD sensors are discussed 
in Sec.~\ref{sec:sensors}; the experimental setups are described in 
Sec.~\ref{sec:setup}; the test beam measurements are reported in 
Sec.~\ref{sec:tb_results}; the beta source setup results and the 
correspondence with probe measurements are presented in 
Sec.~\ref{sec:beta_campaign}; and the conclusions are given 
in Sec.~\ref{sec:conclusion}.

\section{LGAD sensors under study}
\label{sec:sensors}

LGAD sensors are produced by introducing 
an additional layer of p$^+$ material (e.g. Boron) close to the n-p junction
of traditional silicon sensors. This results in a very high electric field 
in the region within a depth of a few micrometers of the junction. 
This region is referred to as the ``gain'' or ``multiplication'' layer. 
The initial signal generated by the MIP ionization is amplified through an 
avalanche process initiated by an electron or a hole accelerating through the 
gain region. The amplified signals maintain fast 
slew rate, resulting in excellent timing characteristics~\cite{Cartiglia201783}. Signals are read out from the n$^+$ 
cathode, and since the bulk material is a high resistivity p-type silicon, 
a shallow uniform p-spray doping is usually implemented to isolate the cathodes. 
To reduce the magnitude of the electric field at the perimeter of each 
signal pad, an additional deep n$^+$ doping region (Junction Terminating 
Extension or JTE) is implemented. JTEs are characteristic feature of LGAD 
sensors, and results in regions of no gain between pads referred to as the 
``inter-pad gap''. 

In this paper we present studies of the new HPK type 3.1 LGAD sensors, 
designed to achieve increased radiation tolerance compared to previous 
LGADs. With a deeper gain implant and generally steeper onset of gain as a 
function of bias voltage, the HPK 3.1 LGADs can be operated at relatively 
low bias voltage before irradiation, and can rely on progressively 
increased bias voltage to compensate for the gain-loss due to the acceptor 
removal that occurs during irradiation. 

\begin{figure}[htbp!] 
	\centering
	\includegraphics[width=0.6\textwidth]{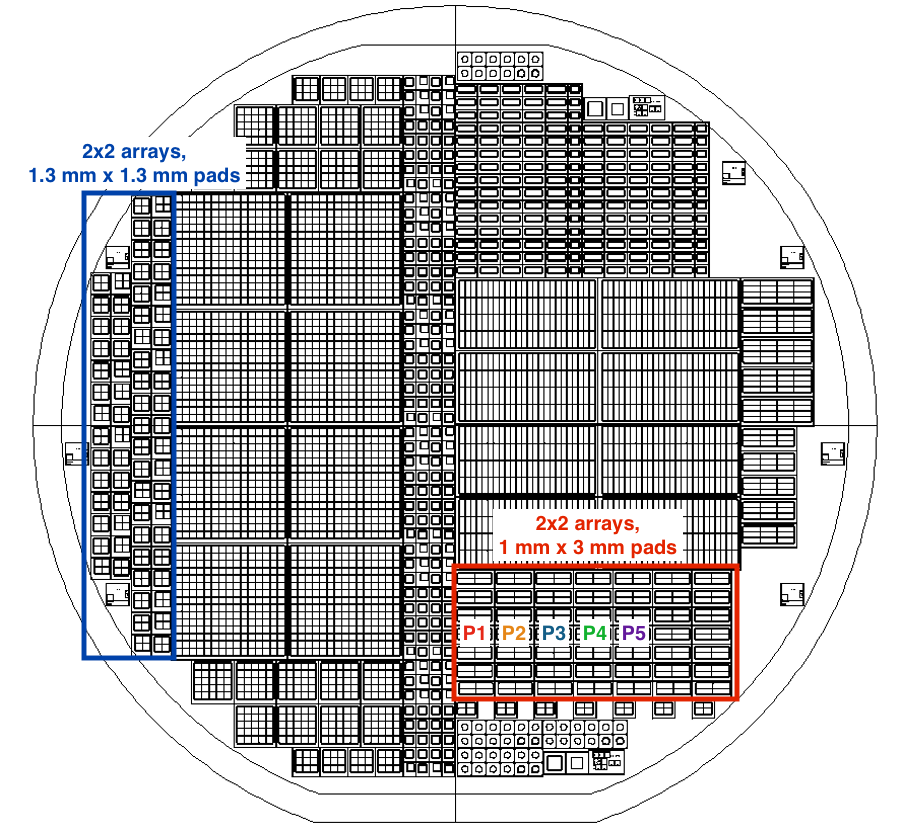}   
	\caption{HPK 3.1 wafer layout. The blue and red boxes indicate the regions of the wafer populated with 2x2 arrays with \SI[product-units = power]{1.3 x 1.3}{\milli \meter} and \SI[product-units = power]{1 x 3}{\milli \meter} pad sizes, respectively. The labels P1-P5 indicate columns of sensors tracked within the population of \SI[product-units = power]{1 x 3}{\milli \meter} sensors. } 
	\label{fig:wafer_layout} 
\end{figure}

The HPK type 3.1 LGADs were produced with a variety of specifications.
The LGADs were produced in 6-inch wafers, comprised of sensors of
different geometries as shown in Fig.~\ref{fig:wafer_layout}. 
The sensors are composed of different grid arrangements of "pads",
which are electrically isolated from each other and read out
separately. The spatial granularity of the pads identify the space points
of the charged particle trajectory. The 
sensors studied in this paper are summarized in Table~\ref{tab:sensors}.

\begin{table}[htbp!]
  \centering

  \begin{tabular}{|l|c|c|c|c|}
  \hline
  Sensor label & Pad arrangement & Fully metalized & Interpad gap & Pad size \\ \hline
   P1 \SI{50}{\micro\m}          & 2$\times$2  & No &   \SI{50}{\micro\m}  & \SI[product-units = power]{1 x 3}{\milli \meter}\\
   P2 \SI{95}{\micro\m}          & 2$\times$2  & No & \SI{95}{\micro\m}    & \SI[product-units = power]{1 x 3}{\milli \meter}\\
   P2 \SI{95}{\micro\m} (metal)          & 2$\times$2  & Yes & \SI{95}{\micro\m}    & \SI[product-units = power]{1 x 3}{\milli \meter}\\
   P2 \SI{50}{\micro\m}          & 2$\times$2  & No & \SI{50}{\micro\m}    & \SI[product-units = power]{1 x 3}{\milli \meter}\\
   P3 \SI{95}{\micro\m}          & 2$\times$2  & No & \SI{95}{\micro\m}    & \SI[product-units = power]{1 x 3}{\milli \meter}\\
   P3 \SI{95}{\micro\m} (metal)          & 2$\times$2  & Yes & \SI{95}{\micro\m}    & \SI[product-units = power]{1 x 3}{\milli \meter}\\
   P3 \SI{50}{\micro\m}          & 2$\times$2  & No & \SI{50}{\micro\m}    & \SI[product-units = power]{1 x 3}{\milli \meter}\\
   P4 \SI{95}{\micro\m}          & 2$\times$2  & No & \SI{95}{\micro\m}    & \SI[product-units = power]{1 x 3}{\milli \meter}\\
   P4 \SI{95}{\micro\m} (metal)          & 2$\times$2  & Yes & \SI{95}{\micro\m}    & \SI[product-units = power]{1 x 3}{\milli \meter}\\
   P4 \SI{50}{\micro\m}          & 2$\times$2  & No & \SI{50}{\micro\m}    & \SI[product-units = power]{1 x 3}{\milli \meter}\\
   P5 \SI{95}{\micro\m} (metal)          & 2$\times$2  & Yes & \SI{95}{\micro\m}    & \SI[product-units = power]{1 x 3}{\milli \meter}\\ \hline
   \SI{90}{\micro\m} \#1         & 2$\times$2  & No & \SI{90}{\micro\m}    & \SI[product-units = power]{1.3 x 1.3}{\milli \meter}\\
   \SI{90}{\micro\m} \#2         & 2$\times$2  & No & \SI{90}{\micro\m}    & \SI[product-units = power]{1.3 x 1.3}{\milli \meter}\\
   \SI{90}{\micro\m} (metal) \#1         & 2$\times$2  & Yes & \SI{90}{\micro\m}    & \SI[product-units = power]{1.3 x 1.3}{\milli \meter}\\
   \SI{90}{\micro\m} (metal) \#2         & 2$\times$2  & Yes & \SI{90}{\micro\m}    & \SI[product-units = power]{1.3 x 1.3}{\milli \meter}\\
   \SI{90}{\micro\m} (metal) \#3         & 2$\times$2  & Yes & \SI{90}{\micro\m}    & \SI[product-units = power]{1.3 x 1.3}{\milli \meter}\\
   \SI{90}{\micro\m} (metal) \#4         & 2$\times$2  & Yes & \SI{90}{\micro\m}    & \SI[product-units = power]{1.3 x 1.3}{\milli \meter}\\
   \SI{90}{\micro\m} (metal) \#5         & 2$\times$2  & Yes & \SI{90}{\micro\m}    & \SI[product-units = power]{1.3 x 1.3}{\milli \meter}\\
   \SI{50}{\micro\m} (metal) \#1         & 2$\times$2  & Yes & \SI{50}{\micro\m}    & \SI[product-units = power]{1.3 x 1.3}{\milli \meter}\\
   \SI{50}{\micro\m} (metal) \#2         & 2$\times$2  & Yes & \SI{50}{\micro\m}    & \SI[product-units = power]{1.3 x 1.3}{\milli \meter}\\
   \SI{30}{\micro\m} (metal) \#1         & 2$\times$2  & Yes & \SI{30}{\micro\m}    & \SI[product-units = power]{1.3 x 1.3}{\milli \meter}\\
   \SI{30}{\micro\m} (metal) \#2         & 2$\times$2  & Yes & \SI{30}{\micro\m}    & \SI[product-units = power]{1.3 x 1.3}{\milli \meter}\\ \hline
   4x4 \SI{95}{\micro\m}        & 4$\times$4   & No &  \SI{95}{\micro\m}  & \SI[product-units = power]{1 x 3}{\milli \meter}\\
   \SI{30}{\micro\m}        & 2$\times$2   & No &  \SI{30}{\micro\m}    & \SI[product-units = power]{1 x 3}{\milli \meter}\\
   \SI{50}{\micro\m}        & 2$\times$2   & No &  \SI{50}{\micro\m}    & \SI[product-units = power]{1 x 3}{\milli \meter}\\
   \SI{70}{\micro\m}        & 2$\times$2   & No &  \SI{70}{\micro\m}    & \SI[product-units = power]{1 x 3}{\milli \meter}\\
   \hline
  \end{tabular}
  \caption{Summary of all the sensors included in the studies described in this paper. The first two blocks list the sensors with pad size \SI[product-units = power]{1 x 3}{\milli \meter} and \SI[product-units = power]{1.3 x 1.3}{\milli \meter} used in the beta source measurements described in Section~\ref{sec:beta_campaign}. The third block lists the sensors used in the test beam measurements described in Section~\ref{sec:tb_results}.} 
  
  \label{tab:sensors}
  \end{table}


Each sensor has a variety of inter-pad separation ranging between \SI{30}{\micro\m} to \SI{100}{\micro\m}. Some sensors were produced 
with complete metalization coverage and some with mostly 
non-metalized open surfaces. Previous studies with metalized and non-metalized LGADs~\cite{ApresyanLGAD} have observed differences in signal propagation time and radiation hardness.
This paper presents a characterization campaign of sensors spanning the full range of the parameters described above. 

The HPK type 3.1 wafer layout is shown in Fig.~\ref{fig:wafer_layout}. Each wafer was split in half, 
with multi-pad sensor arrays of pad size \SI[product-units = power]{1.3 x 1.3}{\milli \meter} on the left, 
and \SI[product-units = power]{1 x 3}{\milli \meter} on the right. The CMS and ATLAS timing 
detectors will employ LGADs with a pad size of \SI[product-units = power]{1.3 x 1.3}{\milli \meter}.

\begin{figure}[htbp!] 
	\centering
	\includegraphics[width=0.6\textwidth]{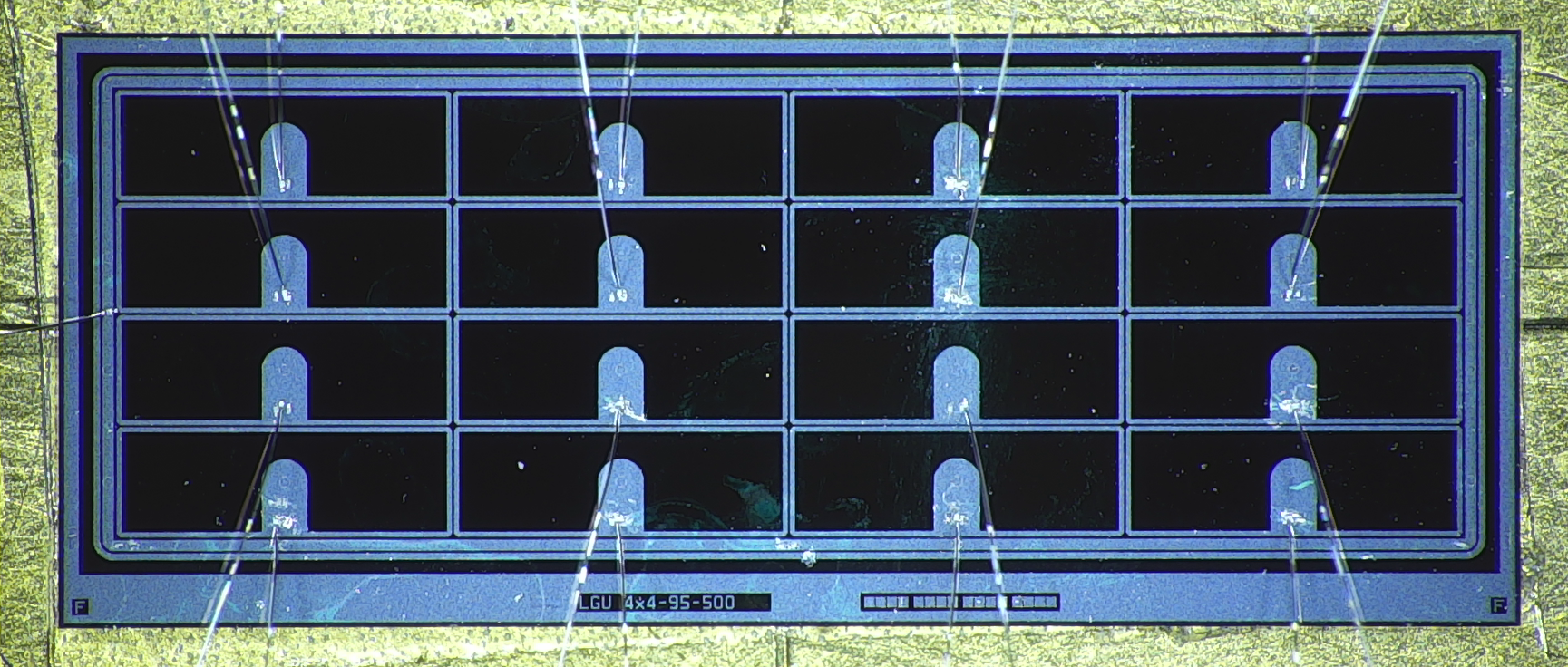}   
	\caption{HPK type 3.1 4x4 multi-pad sensor with pad size \SI[product-units = power]{1 x 3}{\milli \meter}) and a non-metalized surface.}
	\label{fig:array_photo} 
\end{figure}

\begin{figure}[htbp!] 
	\centering
		\includegraphics[width=0.21\textwidth]{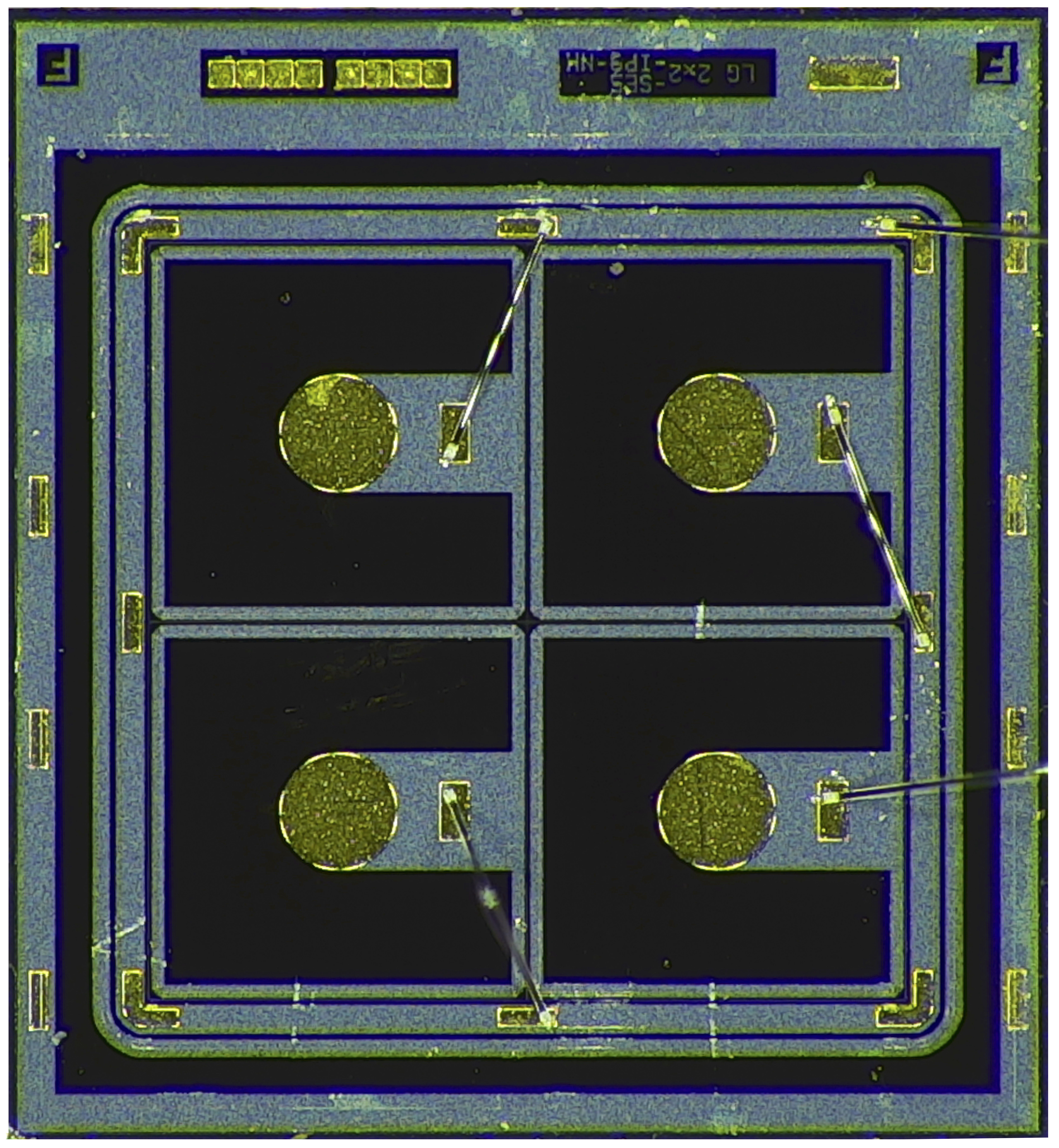}   
	\includegraphics[width=0.46\textwidth]{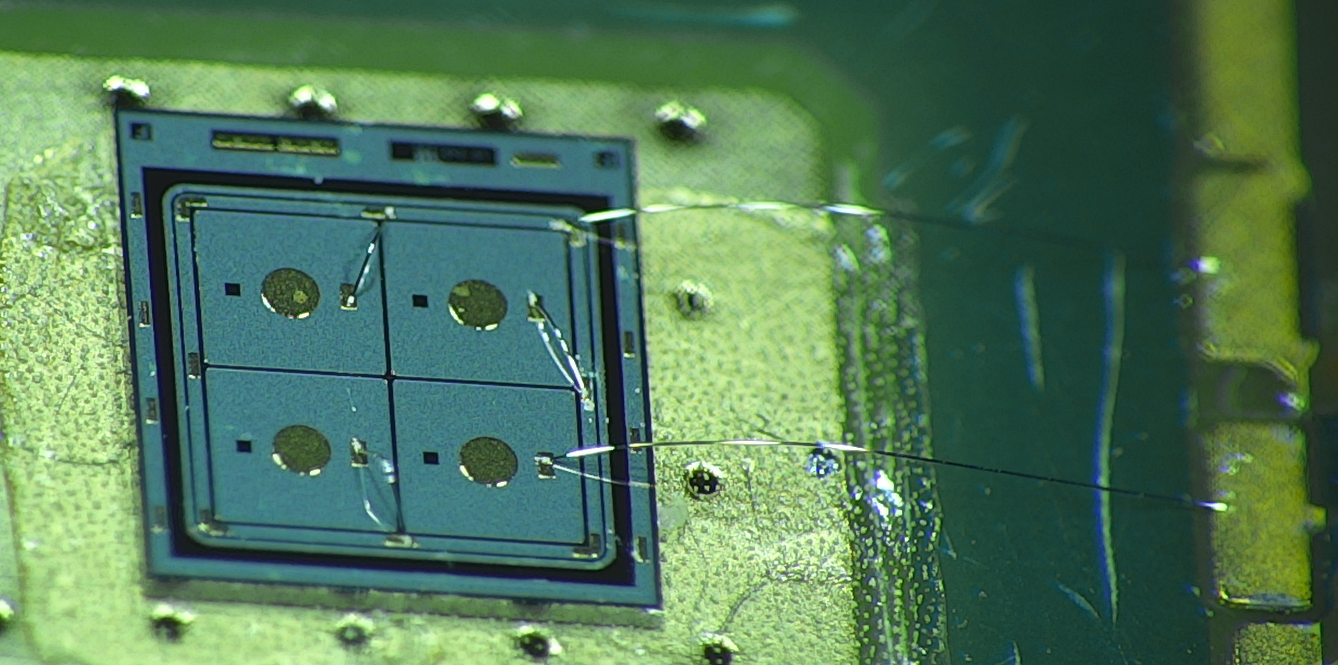}   
	\caption{HPK type 3.1 2x2 multi-pad sensor of pad size \SI[product-units = power]{1.3 x 1.3}{\milli \meter}) without metalized surface (left) and with fully metalized surface (right).}
	\label{fig:2x2_photo} 
\end{figure}

The test beam measurements described in this paper focus on a detailed study of a non-metalized 4 x 4 multi-pad sensor
with \SI[product-units = power]{1 x 3}{\milli \meter} pad size and \SI{95}{\micro \m} inter-pad gap. A photo of 
this sensor is shown in Fig.~\ref{fig:array_photo}. Additionally, several 
similar 2 x 2 multi-pad sensors were studied spanning a variation in the inter-pad spacing 
from \SI{30}{\micro\m} to \SI{95}{\micro\m}. Example photos of these sensors can be seen in Fig.~\ref{fig:2x2_photo}. Since these sensors are typically studied on single channel readout boards (described in Sec.~\ref{ssec:boards}), the three unread pads are grounded via connection to the sensor guard ring.

The beta source telescope, described in Section~\ref{ssec:beta_setup} allows for 
testing a significantly higher volume of sensors thanks to much greater availability and 
ease of access than the test beam. This paper describes a campaign that characterized a sample of 22 sensors and 
documented the variation of the performance metrics using signals from ionizing particles. The campaign 
used a set of 2 x 2 multi-pad sensors with pad sizes of (\SI[product-units = power]{1.3 x 1.3}{\milli \meter} 
and \SI[product-units = power]{1 x 3}{\milli \meter}), spanning the full range of inter-pad gaps, 
and including both metalized and non-metalized sensors. The design parameters of all 22 sensors 
can be found in Table~\ref{tab:sensors}. Prior to the beta source telescope measurements, all of 
these sensors were characterized with capacitance-voltage (CV) measurements using a probe station in the Torino UFSD lab. 
This characterization campaign allows us to extract the precise relationship between the probe station and 
particle measurements. 
For the sensors with \SI[product-units = power]{1 x 3}{\milli \meter} pad size, it was possible to trace 
the provenance to the approximate position on the wafer itself, as indicated in Fig.~\ref{fig:wafer_layout}.

\section{Experimental Setup} 
\label{sec:setup}

\subsection{LGAD readout boards}
\label{ssec:boards}

The test beam and beta source measurements make use of two specialized readout boards: the single-channel UCSC board, optimized for time resolution measurements; and the 16-channel FNAL board, which enables simultaneous readout of many more sensor pads.

The UCSC board is based on a single amplification stage which provides a very clean single-channel measurement. A \SI{470}{\ohm} transimpedance amplifier with an analog bandwidth of \SI{1.6}{\giga \hertz} is used, with very low noise, due to extensive shielding and isolation~\cite{Cartiglia201783}.
In all measurements described in this paper, the UCSC board is supplemented with an additional stage of amplification provided by a Mini-Circuits GALI-52+ evaluation board, resulting in a total transimpedance of approximately \SI{5}{\kilo \ohm}. Several copies of the UCSC board (v1.4) were used and found to have excellent inter-board uniformity, as described in Section \ref{ssec:board_cal}.
The 16-channel FNAL board enables readout of large multi-pad LGAD sensors with sizes up to $26.5\times11.5~\rm{mm}^2$. Each readout channel is amplified by two amplifier stages based on the Mini-Circuits GALI-66+ integrated circuit. In this particular configuration, the amplifiers used a \SI{25}{\ohm} input impedance, for a total transimpedance of approximately \SI{5}{\kilo \ohm} and a bandwidth of \SI{1}{\giga \hertz}. The amplifier chain of each readout channel is found to have uniform gain with approximately 10\% variation from channel to channel. 

In general, the UCSC board is able to provide a better time resolution measurement for a single sensor pad, while the 16-channel board affords the ability to study multiple sensor pads with a time resolution that is slightly degraded.

\subsection{Experimental chambers and data acquisition}
\label{ssec:chamber}

For all test beam, beta source, and laser measurements, the LGADs were 
mounted inside similar experimental chambers that provide a controlled 
environment and enable stable, reproducible results. Inside the chambers, 
the sensors and read-out boards were coupled to aluminum cooling blocks 
shown on Figure~\ref{fig:beta_setup}, which are mounted on a remotely 
operated motorized stage. The cooling blocks are machined with a dense 
network of cooling channels that ensure efficient cooling of the mounted 
electronics. A chilled glycol-water solution was circulated through the 
cooling blocks, capable of holding the sensors at constant temperatures 
ranging from \SI{-20}{\celsius} to +\SI{+22}{\celsius}, as verified with 
on-board thermistors. Particularly important for the higher-power 
16-channel board, direct contact with the cooling blocks enables 
significantly more heat dissipation than possible with cooling via contact with cold air alone. All relevant sensor operational parameters were continuously monitored and recorded, including the sensor 
temperature via on-board thermistors, the sensor bias voltage and leakage 
current, and the temperature and humidity of the air. These values were 
subsequently synchronously combined with the signal waveform data. The 
temperature of the readout boards were found to be stable to within 
\SI{0.1}{\celsius}. The relative humidity of the environmental chamber was 
kept to less than $10\%$ by means of a constant flow of nitrogen gas. 

For both the test beam and the beta source measurements, a precise 
reference time stamp is provided by a Photek $240$~micro-channel plate 
(MCP-PMT) detector, placed inside the environmental chambers just 
behind the LGADs~\cite{Photek240}. Cherenkov radiation emitted by protons 
or beta particles traversing the glass window produce photoelectrons that 
generate a large, steeply rising MCP-PMT signal. The time resolution of the 
MCP-PMT response to protons in the test beam was measured to be better than 
\SI{10}{\pico \second} by comparing timestamps produced by two Photek $240$ 
MCP-PMTs aligned in the beamline at once. Due to the lower intensity of 
Cherenkov radiation emitted by beta particles, the MCP-PMT signals in the 
beta setup have smaller amplitudes, resulting in a time resolution of
approximately \SI{15}{\pico \second}, as described in 
Section~\ref{ssec:tb_beta_validation}.

The LGAD and MCP-PMT waveforms were acquired using a Keysight MSOX92004A 
4-channel oscilloscope~\cite{Keysight}, which provides digitized waveforms 
sampled at 20--40 GS/s, and the oscilloscope bandwidth was set 
to \SI{2}{\giga \hertz}.

\subsection{Fermilab Test Beam Facility experiment}
\label{ssec:beam_setup}

The test beam measurements were performed at the FTBF~\cite{FTBF}, which provides a unique opportunity to characterize prototype detectors for collider experiments. The FTBF uses the \SI{120}{\GeV} proton beam from the Fermilab Main Injector accelerator. The FTBF beam is resonantly extracted in a slow spill for each Main Injector cycle delivering a single \SI{4.2}{\s} long spill per minute, tuned to yield approximately 100,000 protons each spill. The primary beam of \SI{120}{\GeV} protons is bunched at \SI{53}{MHz}. The beam size can be tuned to obtain widths from $2$-\SI{3}{\milli \m} up to approximately \SI{1}{\centi \m}. All measurements presented in this paper were taken with such primary beam particles.

\begin{figure}[htbp!]
   \centering

    \includegraphics[width=0.9\textwidth]{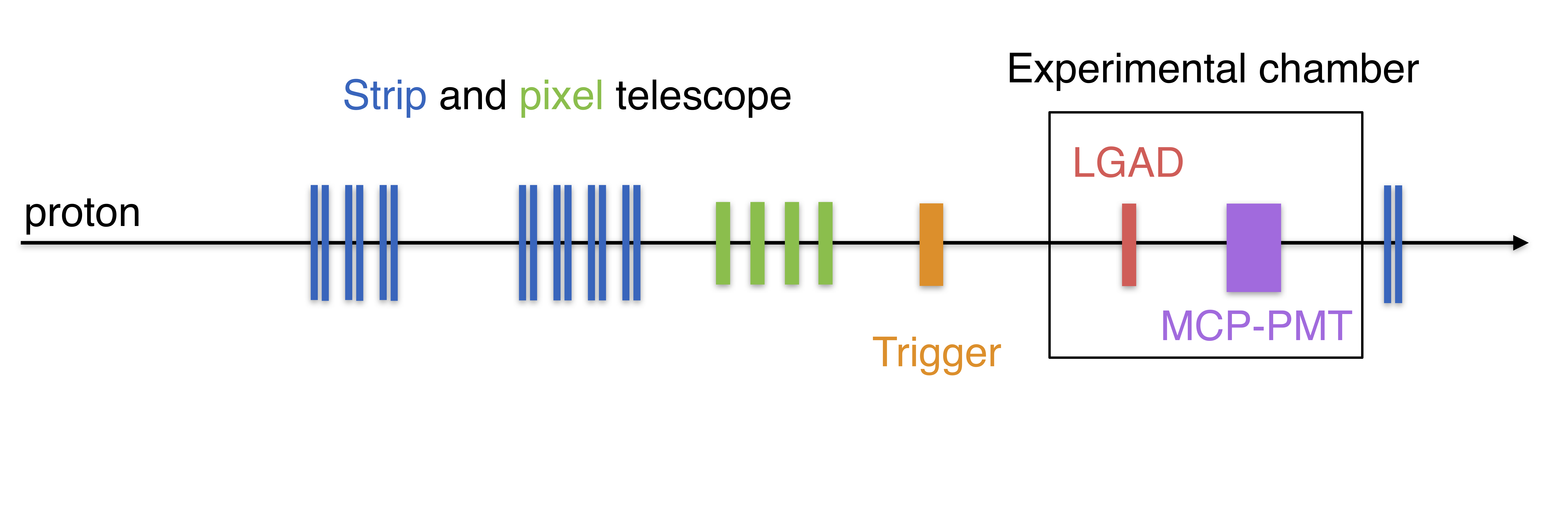}
    
    \includegraphics[width=0.7\textwidth]{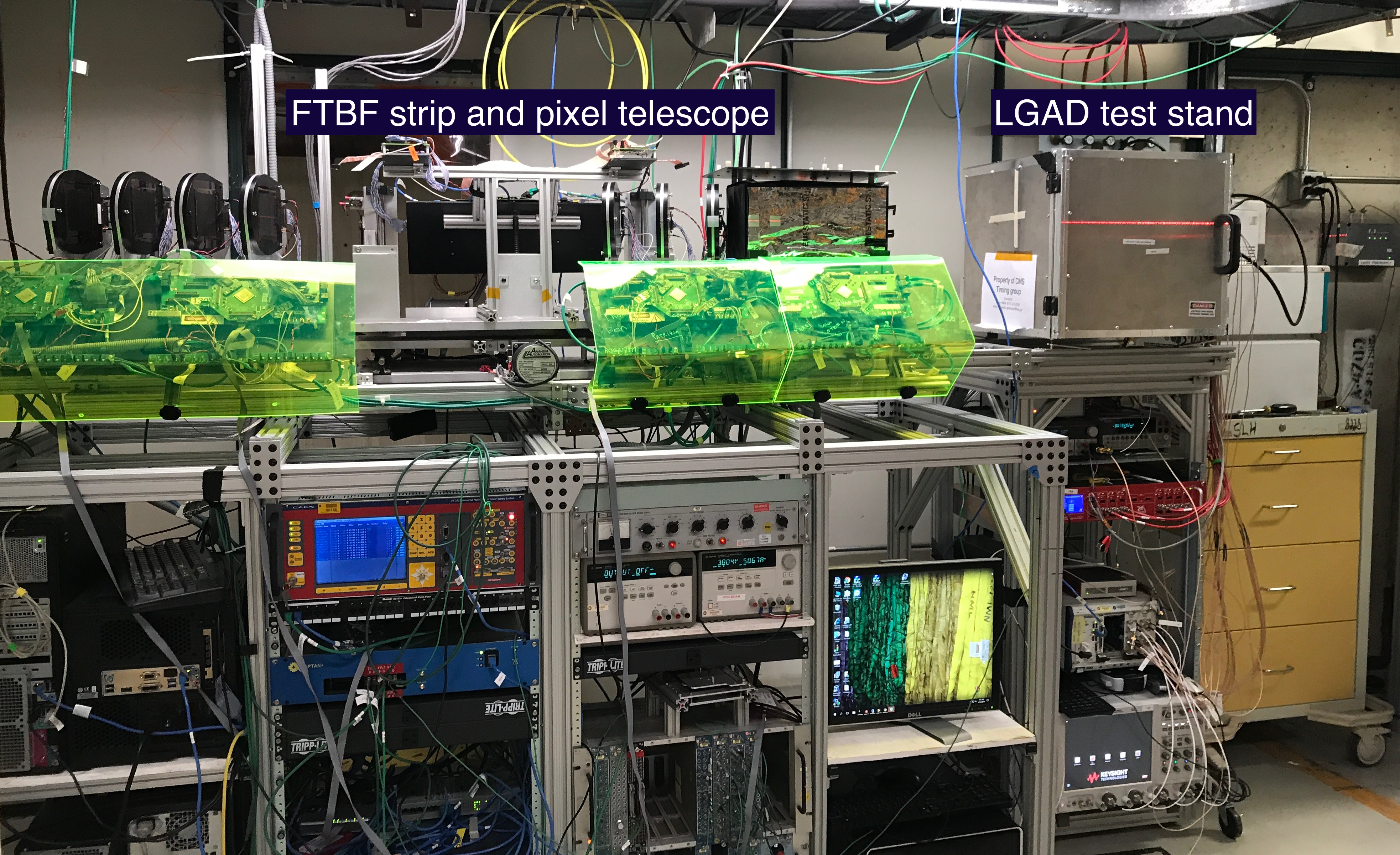}

\caption{A schematic diagram of the test beam setup and FTBF telescope geometry (top) A photo of the experimental setup and the telescope tracker at FTBF.}
\label{fig:TB_setup}
\end{figure}

The FTBF is equipped with a silicon tracking telescope to measure the 
position of each incident proton~\cite{KWAN2016162}. The telescope consists of four pixel 
layers with cell size \SI[product-units = power]{100 x 150}{\micro \meter}, 
and fourteen strip modules with \SI{60}{\micro \meter} pitch, in 
alternating orientation along the x- and y- axes of the plane orthogonal 
to the beam axis. The LGAD chamber was placed approximately \SI{2}{\meter} 
downstream from the center of the telescope. To detect and reject protons 
that scatter in any material along the beam line, two of the fourteen strip 
layers are located immediately downstream of the LGAD chamber. 
During this data-taking period, the resolution of the telescope spatial 
measurement at the LGAD position was approximately \SI{50}{\micro \meter}, 
somewhat degraded with respect to its nominal resolution of 
10--15\si{\micro\meter} due to the long extrapolation from the 
telescope and material along the beamline. The FTBF tracker setup and the 
LGAD environmental chamber can be seen in Fig.~\ref{fig:TB_setup}.

The telescope data acquisition hardware is based on the CAPTAN 
(Compact And Programmable daTa Acquisition Node) system developed at 
Fermilab. The CAPTAN is a flexible and versatile data acquisition system 
designed to meet the readout and control demands of a variety of pixel and 
strip detectors for high energy physics applications~\cite{4775101}. 

The trigger signal to both the telescope and the oscilloscope originates in 
an independent scintillator coupled to a photomultiplier tube. The 
telescope and oscilloscope data are merged offline by matching trigger 
counters from each system. 

The Keysight MSOX92004 oscilloscope's extremely deep memory and segmented 
acquisition mode is particularly well suited for the FTBF beam structure, 
allowing a burst of $50,000$ events to be acquired during each 
\SI{4.2}{\second} spill and written to disk during the longer inter-spill 
period. To efficiently read the multi-pad sensors using only three channels 
on the oscilloscope, multiple readout channels are connected to a high performance 
\SI{20}{\giga \hertz} RF switch~\cite{multiplexer}. Then, the signals
from different sensor pads can be remotely selected as input to the oscilloscope
for readout without interrupting the beam.

The data acquisition and reconstruction process is managed by an automated 
software framework known as JARVIS~\cite{jarvis}. This framework 
synchronizes the operation of the oscilloscope, telescope, and any other 
instruments with the accelerator. After each run, JARVIS uploads all of the 
run parameters as well as any environmental monitoring to a run database 
based on the AirTable framework. JARVIS then manages the reconstruction and 
merging of the various datastreams using distributed computing resources at 
the Fermilab LHC Physics Center (LPC).

\subsection{Fermilab beta source experiment}
\label{ssec:beta_setup}

\begin{figure}[htbp!]
   \centering
    \includegraphics[width=0.54\textwidth]{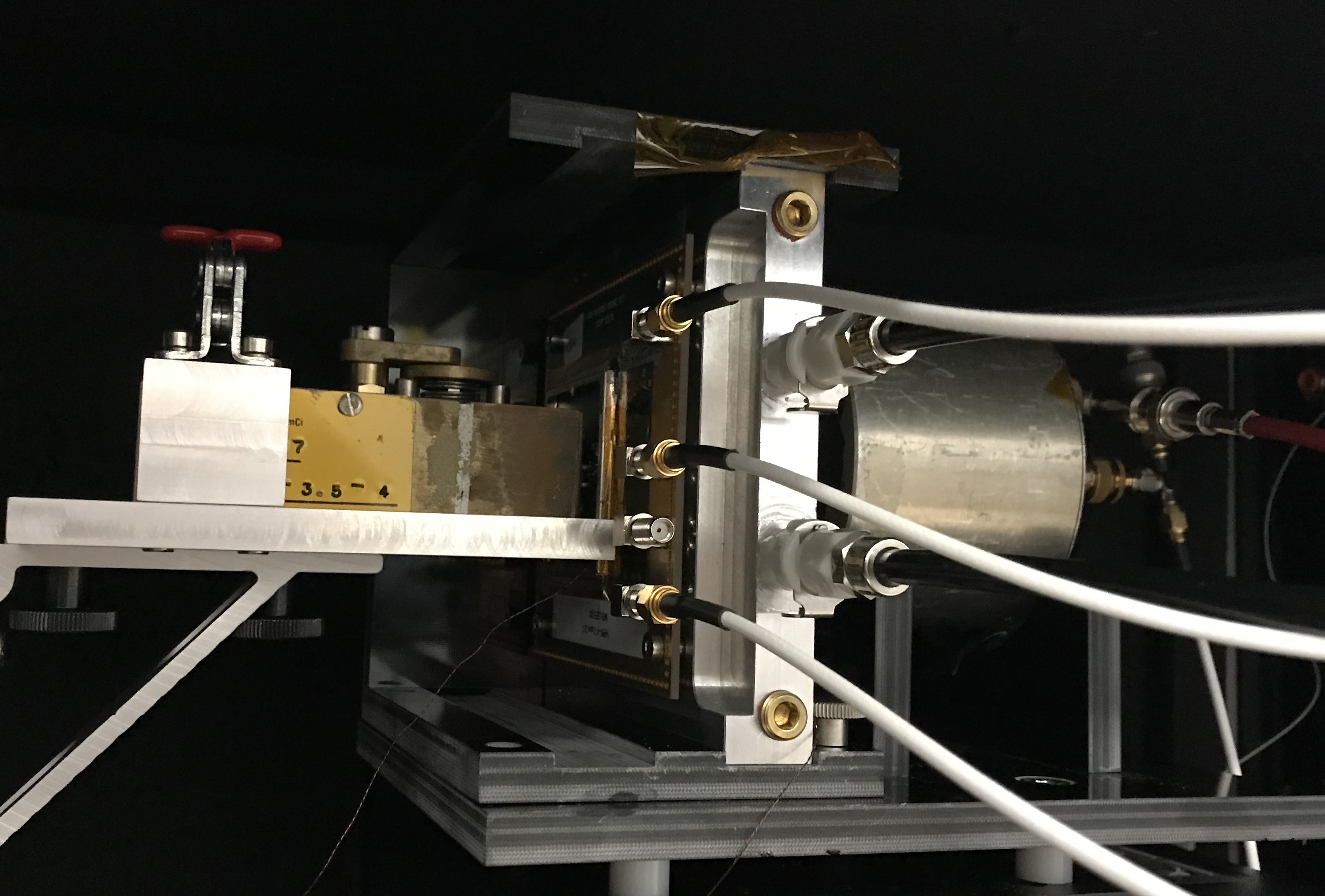}
    \includegraphics[width=0.44\textwidth]{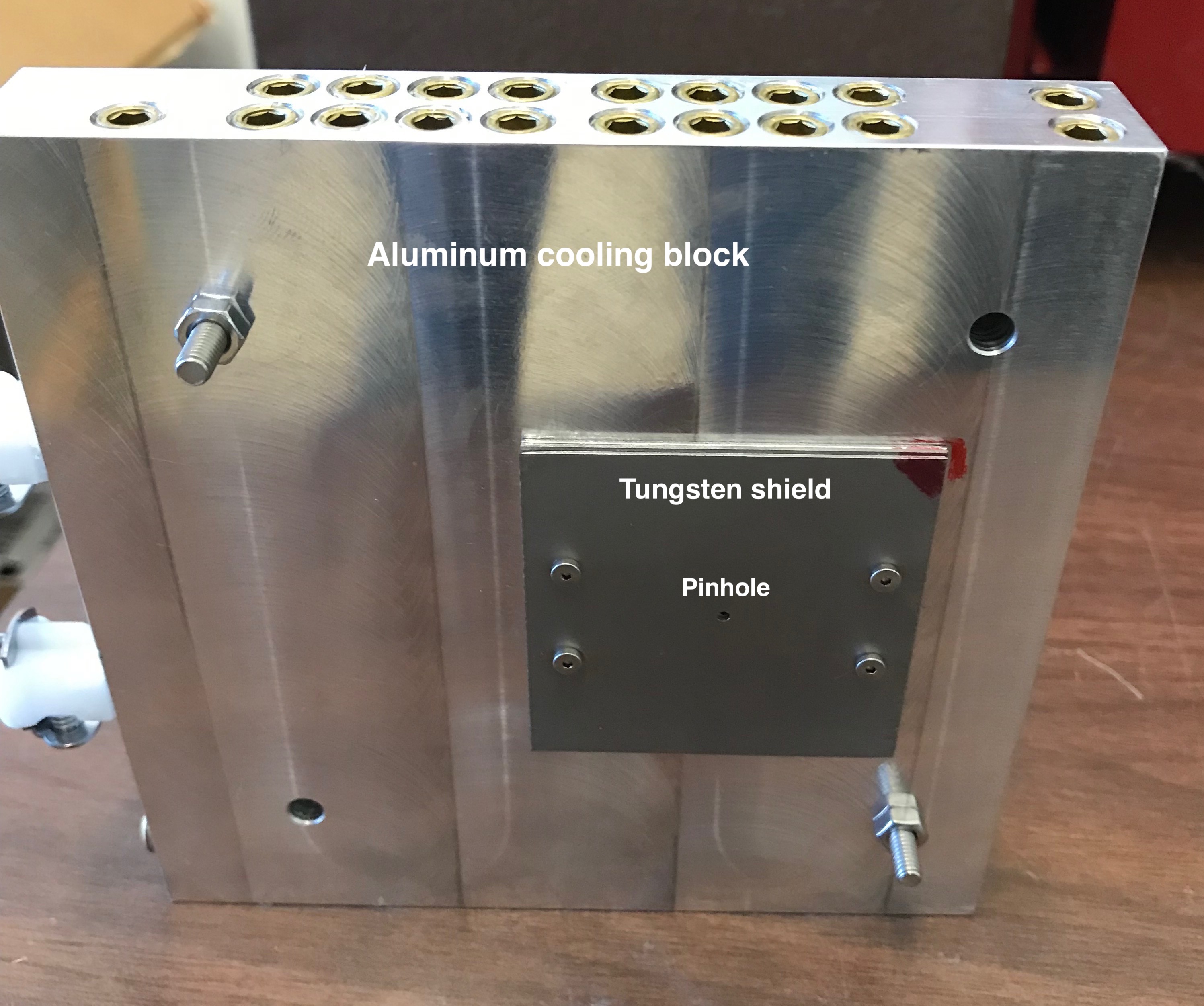}
\caption{The FNAL beta source setup, showing the beta gun, the LGAD readout board mounted on cooling block, and the MCP-PMT (left). Backside of the cooling block with tungsten pinhole attached (right).}
\label{fig:beta_setup}
\end{figure}

The beta source measurements were performed using a Ruthenium-106 source with an activity of approximately \SI{1.3}{\milli \curie}, and a typical beta energy of 1--2 MeV. The beta source is stored inside a brass "beta gun" collimator and mounted on a stand facing the sensor. All of the beta source measurements shown in this paper use the UCSC board described in Section~\ref{ssec:boards}, with a \SI{1.3}{\milli\m} hole drilled underneath the sensor to facilitate the passage of the beta rays. To reject events where the beta particle scatters inside the sensor, a tungsten shield with a \SI{1.5}{\milli \m} pinhole is mounted on the backside of the cooling block, about \SI{2}{\centi \m} behind from the sensor. Particles that pass through the pinhole then reach the MCP-PMT, which serves as both the trigger and the timing reference. By selecting beta rays that pass through the LGAD sensor, the pinhole, and the MCP-PMT, the resulting population of events have a uniform path length as well a MIP-like distribution of energy deposition in the sensor, as will be demonstrated in Section~\ref{ssec:tb_beta_validation}. Although the rate of beta rays that leave large, non-MIP charge deposits is significant in general, these events do not reach the MCP-PMT and are efficiently rejected. Photos of the beta source setup can be seen in Fig.~\ref{fig:beta_setup}.

\section{Test beam results}
\label{sec:tb_results}

\begin{figure}[htbp!] 
\centering
\includegraphics[width=0.49\textwidth]{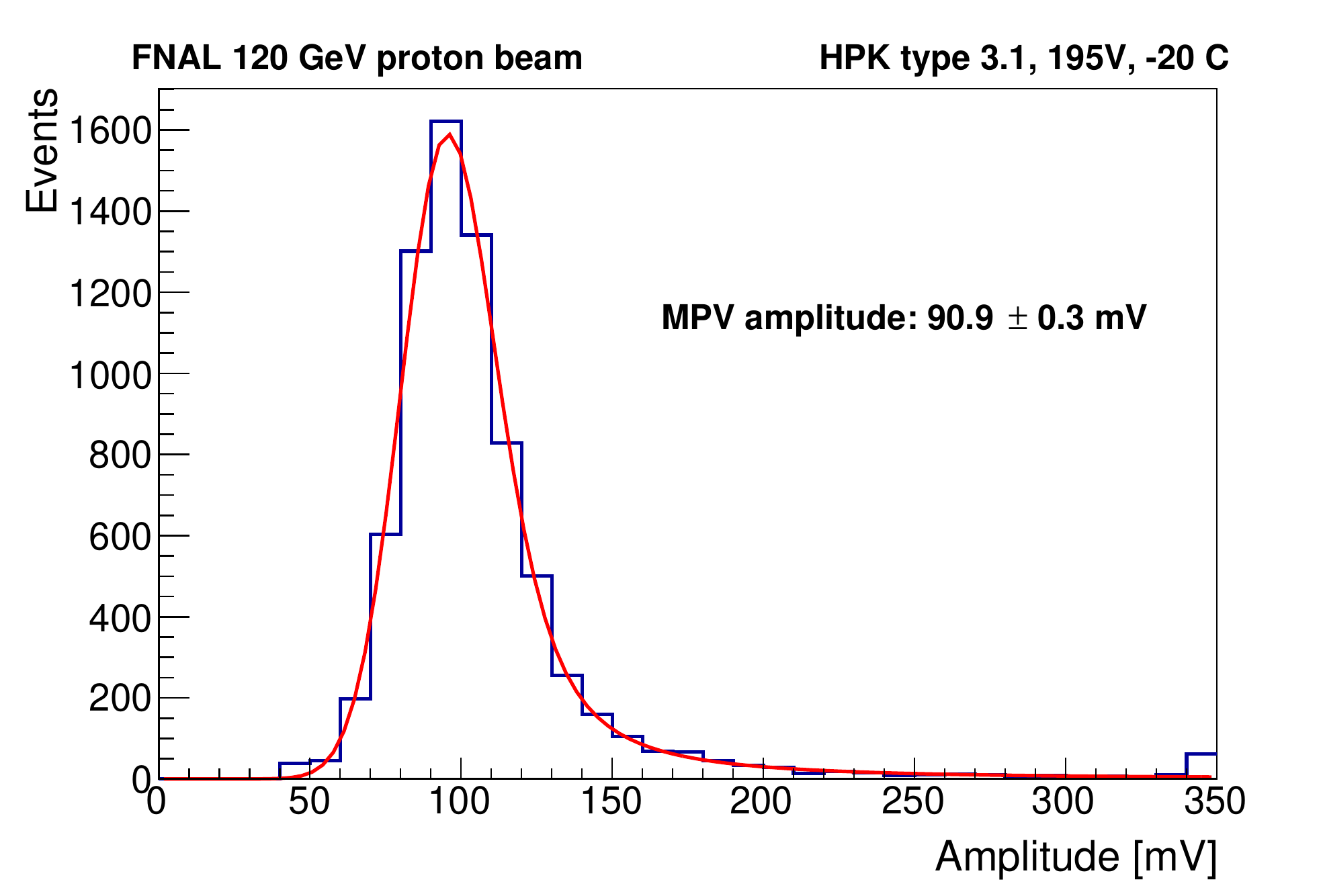} 
\includegraphics[width=0.49\textwidth]{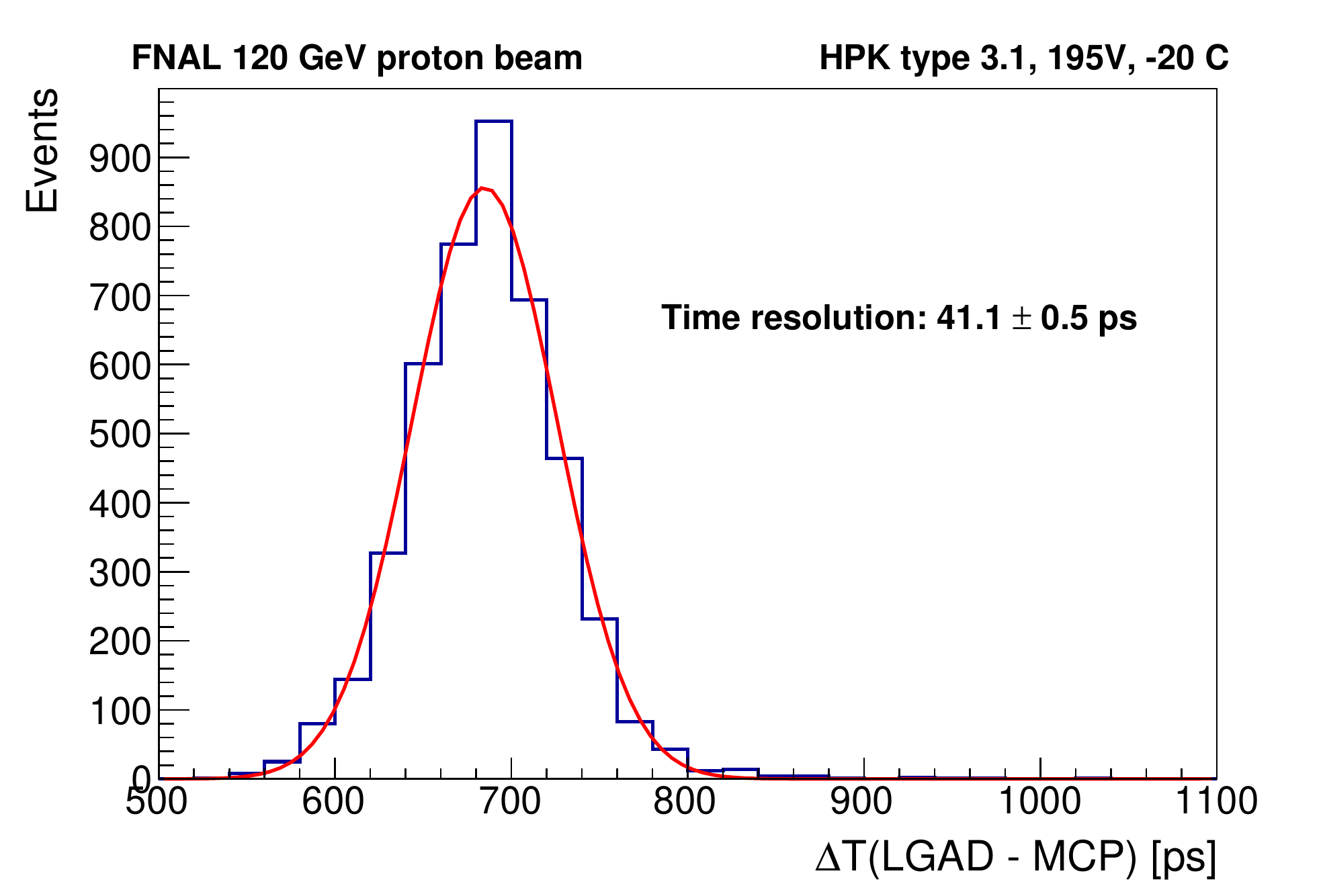}  

\caption{Distributions from a single LGAD pad of the signal amplitude (left) and time of arrival difference with respect to the MCT-PMT (right). The amplitude distribution is fit with a Landau distribution convolved with a Gaussian to extract the most probable value. The time difference distribution is fit with a Gaussian function to extract the time resolution.} 
\label{fig:example_histograms} 
\end{figure}

The HPK 3.1 4x4 multi-pad sensor described in Section~\ref{sec:sensors}, mounted on the 16-channel readout board described in Section~\ref{ssec:boards} was exposed to the beam to study the response of the sensor to minimum ionizing particles. Fig.~\ref{fig:example_histograms} shows example signal amplitude and time of arrival distributions collected from protons passing through a single pad of the array.

We further present studies of the signal response across the entire area of the sensor, using the FTBF tracker to determine the location of the proton in each event. For these 
studies, a good quality track from the FTBF tracker is required, including 
hits in the final strip planes behind the sensor in order to reject protons 
that are deflected in material along the beamline.

\begin{figure}[htbp!] 
\centering
\includegraphics[width=0.75\textwidth]{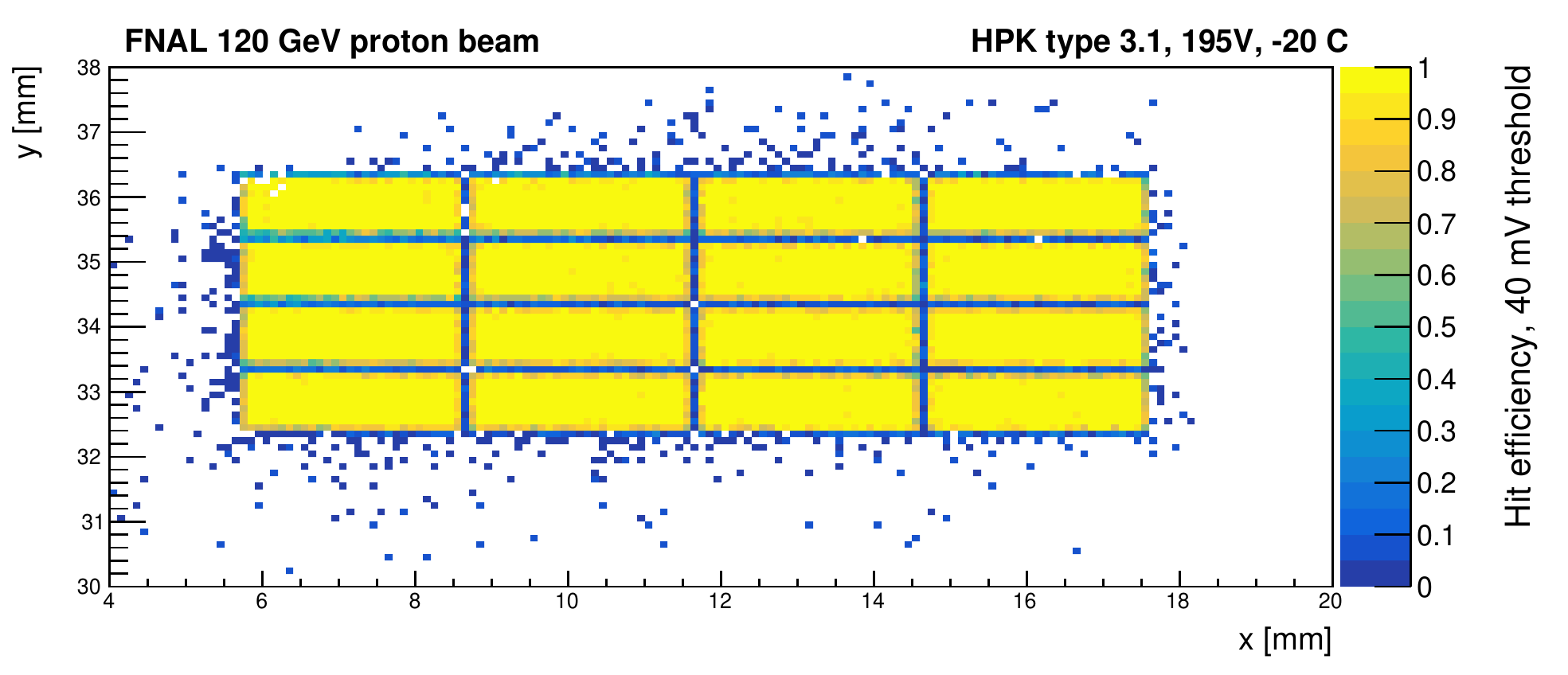}  
\includegraphics[width=0.49\textwidth]{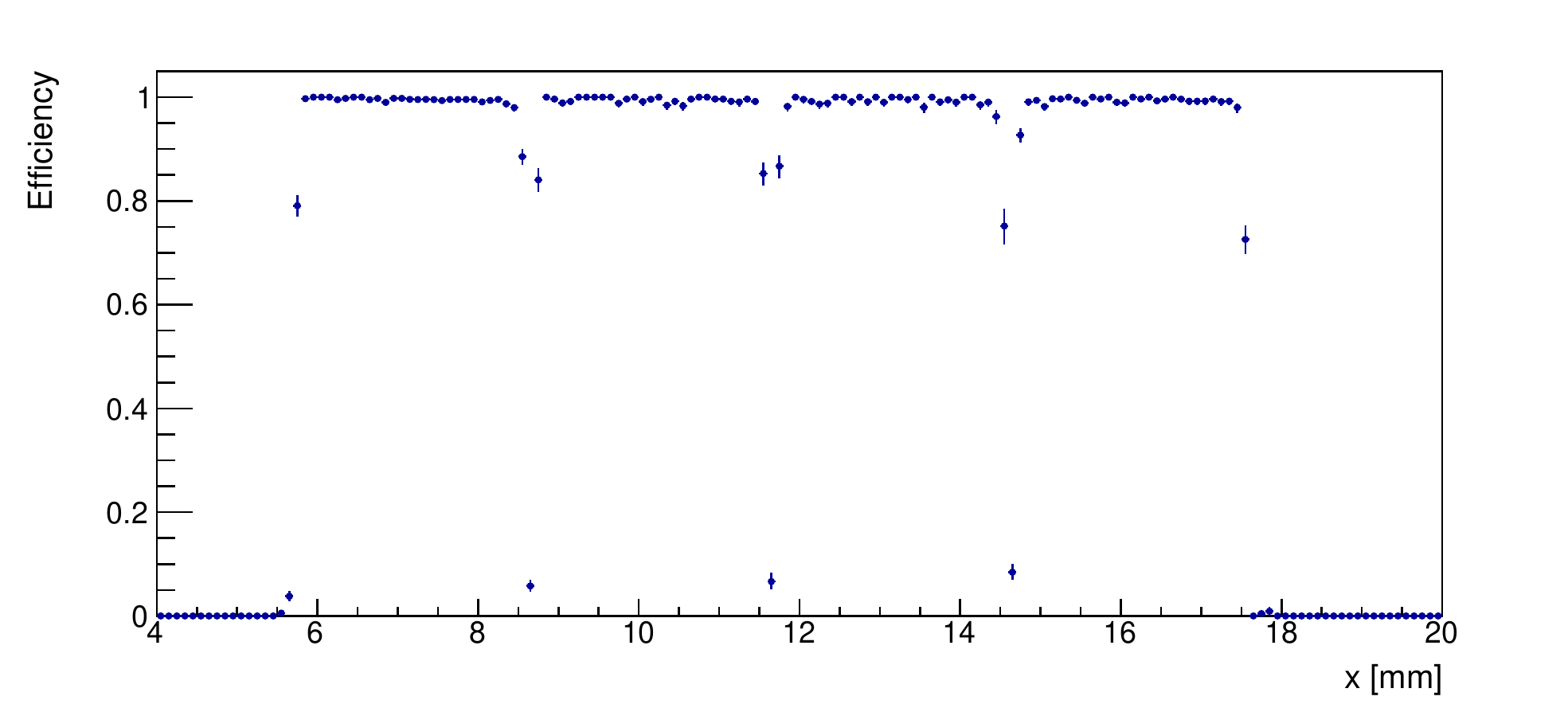}  
\caption{Hit efficiency across surface of the 4x4 sensor for a threshold of \SI{40}{\milli \volt} (top). Projection of the efficiency along the x-axis for the second row of pads, averaging over the region $33.6 <$ y $< 35$ mm (bottom).} 
\label{fig:eff_4x4} 
\end{figure}

The hit efficiency across the surface of the sensor is shown in Fig.~\ref{fig:eff_4x4}. For a given position on the sensor, the efficiency is defined as the fraction of events with a signal amplitude larger than \SI{40}{\milli \volt} out of all events with a high-quality track pointing through that position. This threshold is significantly larger than the noise and corresponds to a selection efficiency greater than 99\% in all active regions. The isolated hits outside the sensor area arise due to events with misreconstructed trajectories or with more than one proton. In this figure, the inter-pad gap regions are clearly visible.

\begin{figure}[htbp!] 
\centering
\includegraphics[width=0.75\textwidth]{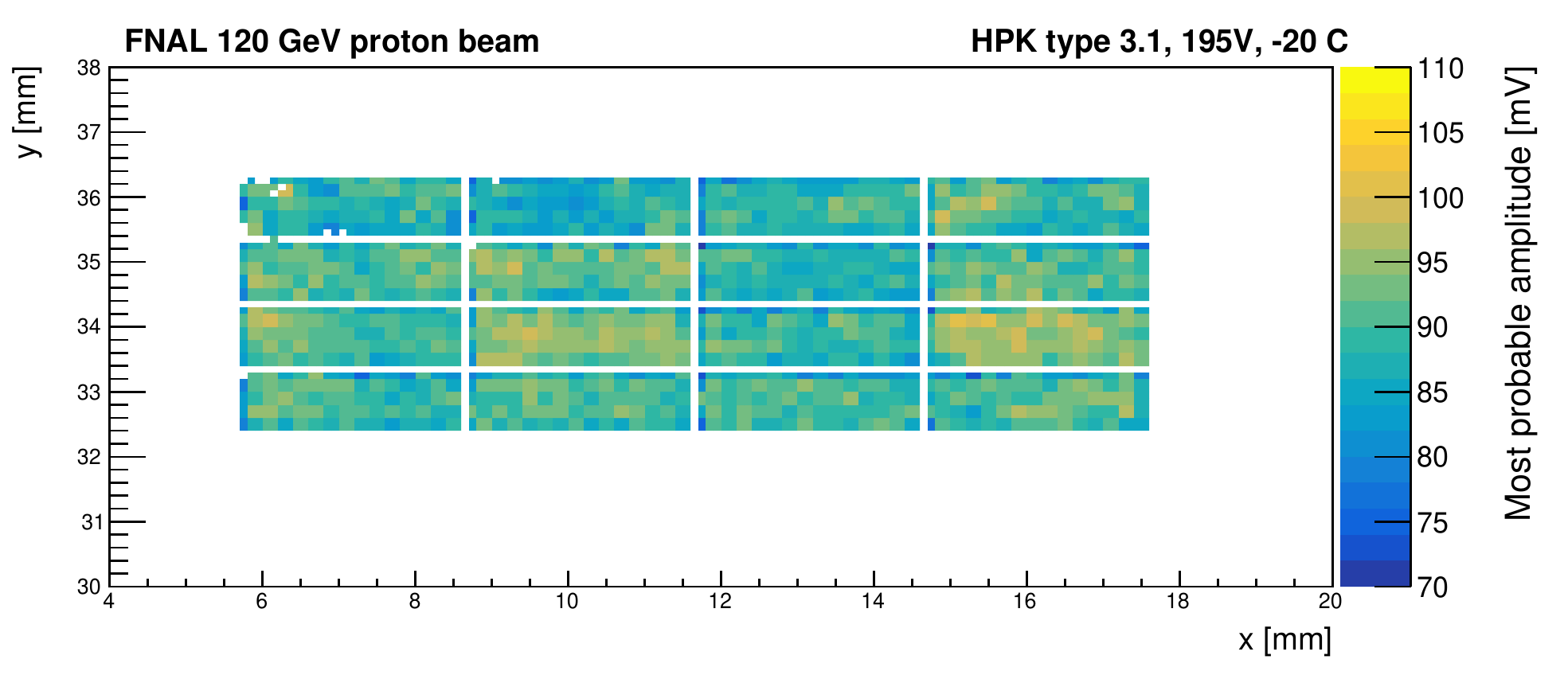}  
\includegraphics[width=0.49\textwidth]{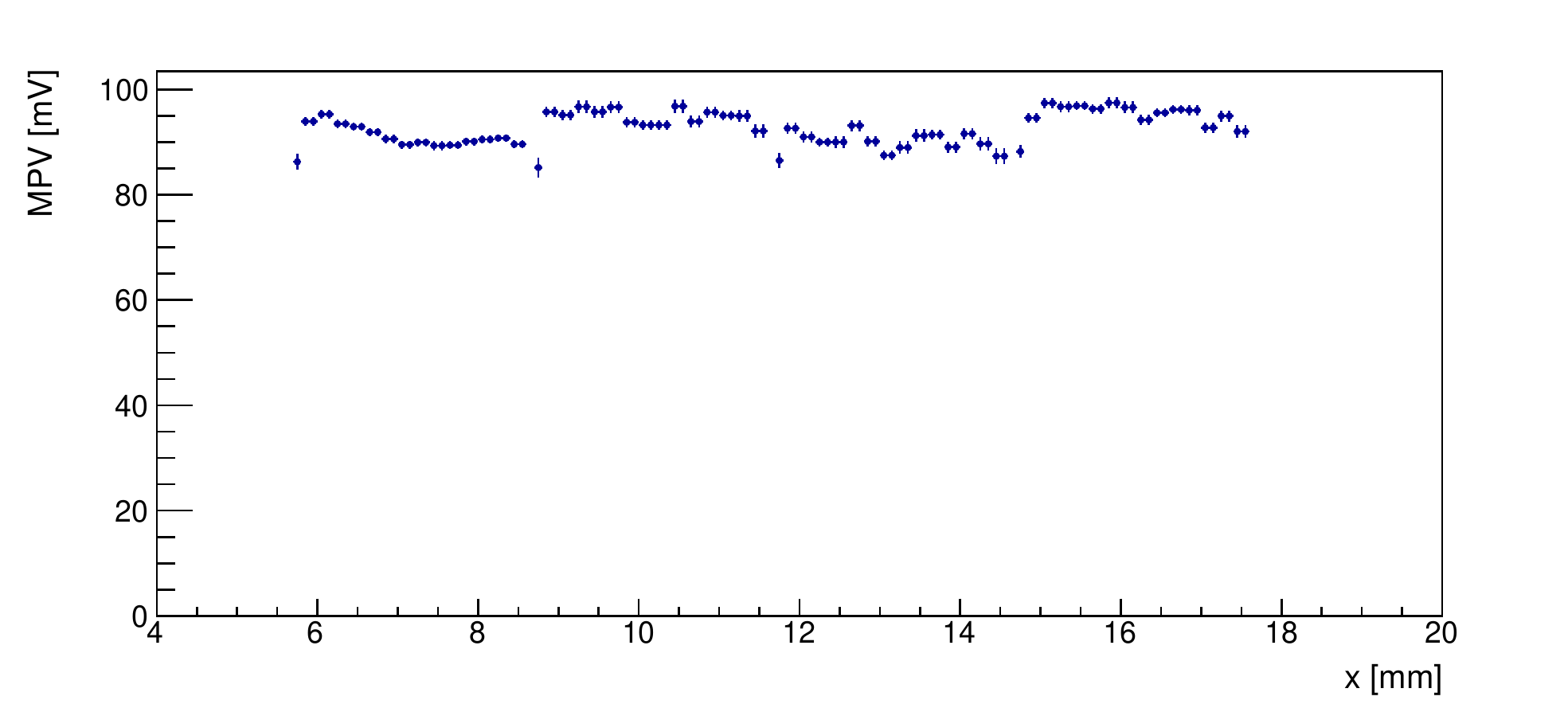}  

\caption{Most probable signal amplitude across the surface of the 4x4 sensor (top). Projection of the most probable amplitude along the x-axis for the second row of pads, averaging over the region $33.6 <$ y $< 35$ mm (bottom).} 
\label{fig:mpv_4x4} 
\end{figure}

Fig.~\ref{fig:mpv_4x4} shows a map of typical signal amplitudes observed across the surface of the sensor. The most probable amplitudes are extracted from a fit to the amplitude distribution using a Landau distribution convolved with a Gaussian distribution. The response is highly uniform, with pad-to-pad variations up to 10\%, consistent with the variations in gain between different amplifiers on the readout board.

\begin{figure}[htbp!] 
\centering
\includegraphics[width=0.75\textwidth]{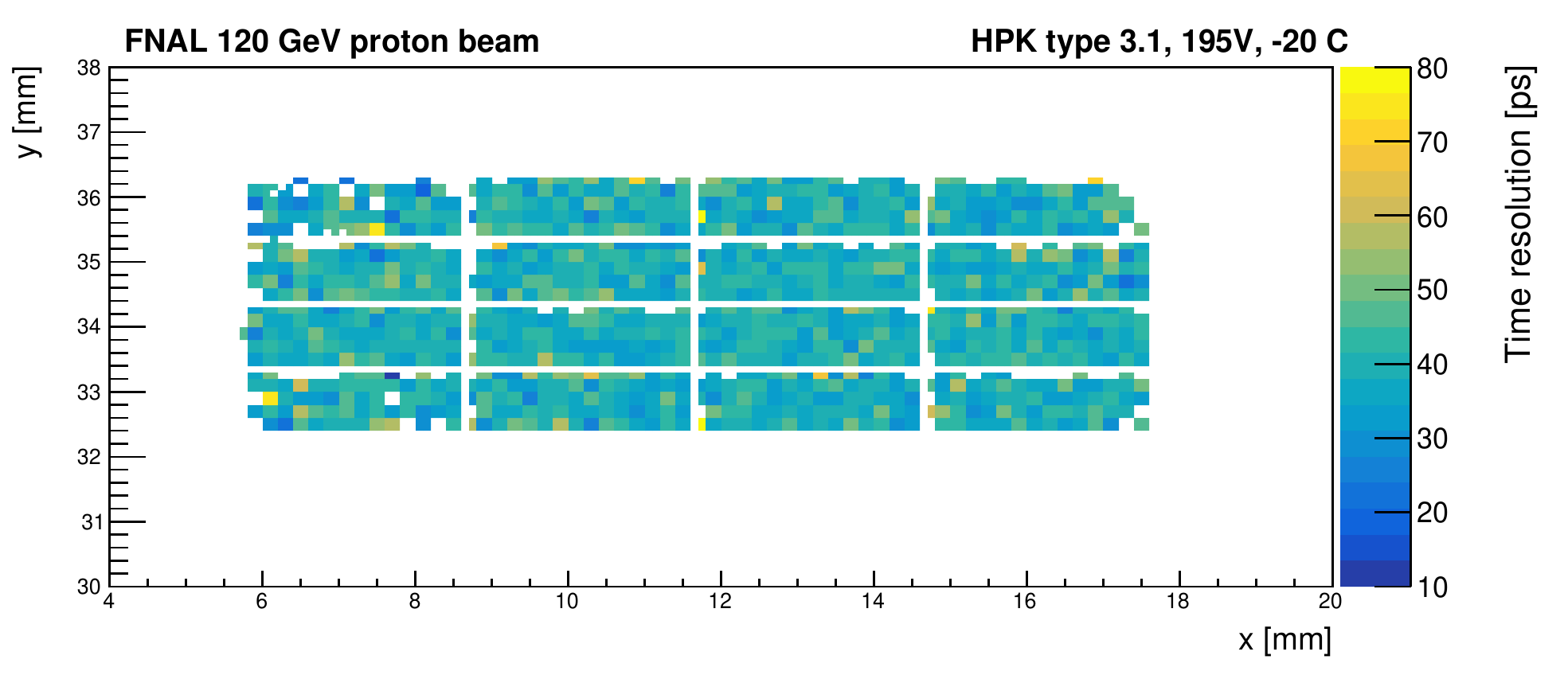}  
\includegraphics[width=0.49\textwidth]{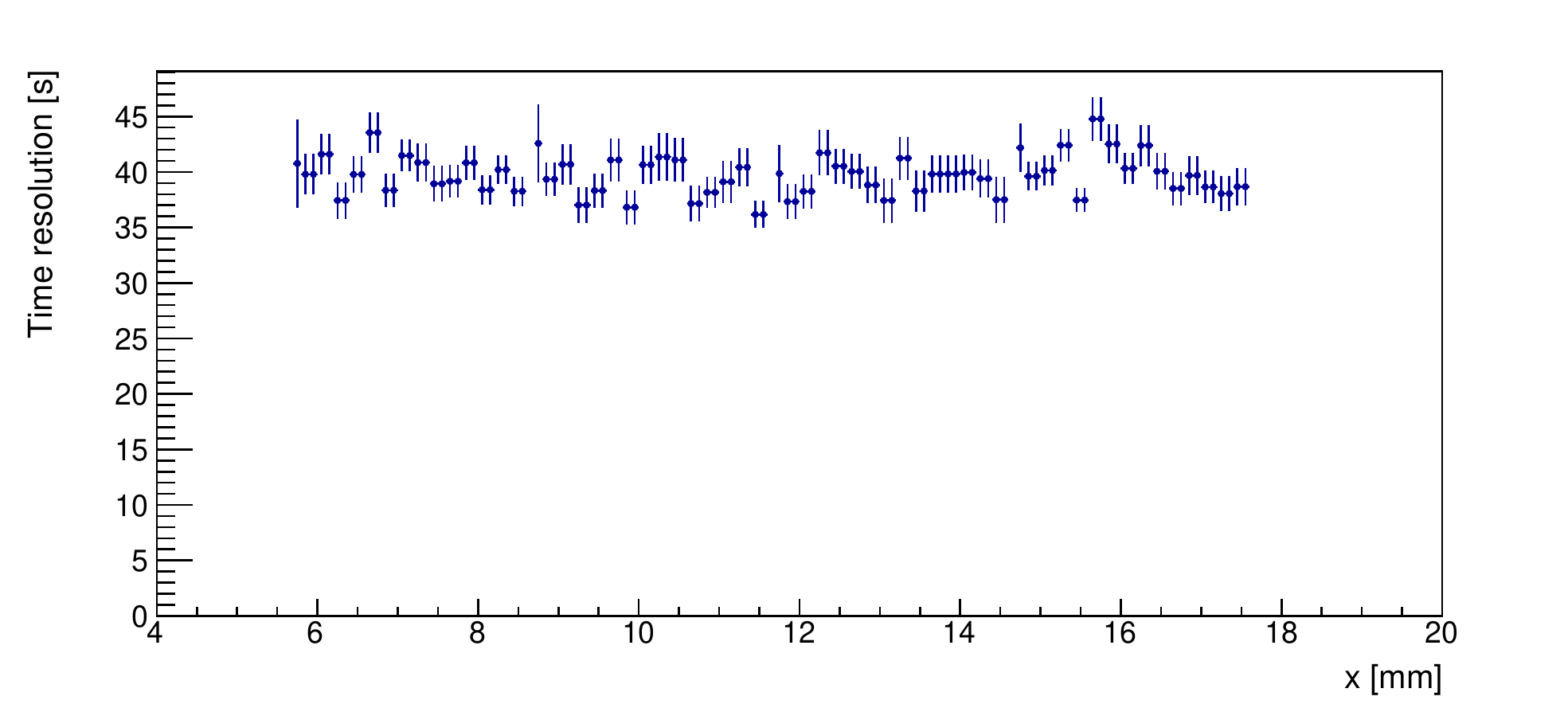}  
 
\caption{Time resolution across surface of the 4x4 sensor (top). Projections of the resolution vs x for the second row of pads, averaging over the region $33.6 <$ y $< 35$ mm (bottom).} 
\label{fig:timeres_4x4} 
\end{figure}

Fig.~\ref{fig:timeres_4x4} shows a map of time resolution across the sensor. The time resolution is extracted as the width of a gaussian fit to the distribution of time difference between the LGAD and MCP-PMT timestamps for all events with a high-quality track at a given position on the sensor. The timestamps for each event are calculated by applying a 20\% (40\% for MCP-PMT) constant fraction discriminator to a linear fit on the rising edge of each waveform between 10\% and 90\% of the maximum amplitude. The time resolution is observed to be uniform for all of the active areas of the pads. Some bins are left empty due to insufficient event counts for performing a reliable fit to the distribution of time difference.

On the 16-channel FNAL board, the time resolution at high bias voltage reaches \SI{40}{\pico\second}, as shown in Fig.~\ref{fig:timeres_4x4}. However, the same sensor on the lower noise UCSC board reaches a resolution of \SI{30}{\pico\second}. This difference is due to slightly higher noise on the 16-channel board. Combining the information from both measurements, we conclude that the intrinsic sensor resolution is uniformly \SI{30}{\pico\second}.

\begin{figure}[htbp!] 
\centering
\includegraphics[width=0.75\textwidth]{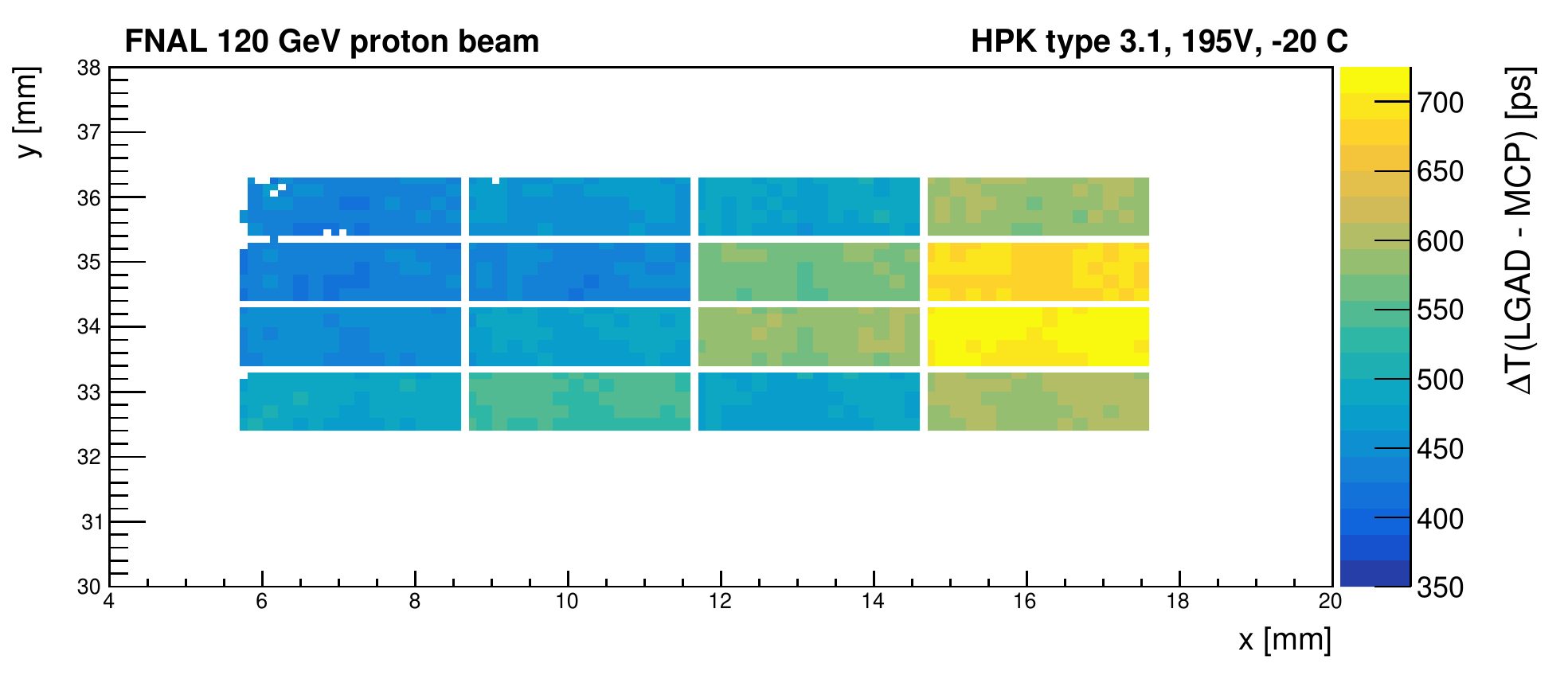}  

\caption{LGAD signal arrival time with respect to the MCP-PMT timestamp across the surface of the 4x4 sensor.} 
\label{fig:timedel_4x4} 
\end{figure}

\begin{figure}[htbp!] 
\centering
\includegraphics[width=0.5\textwidth]{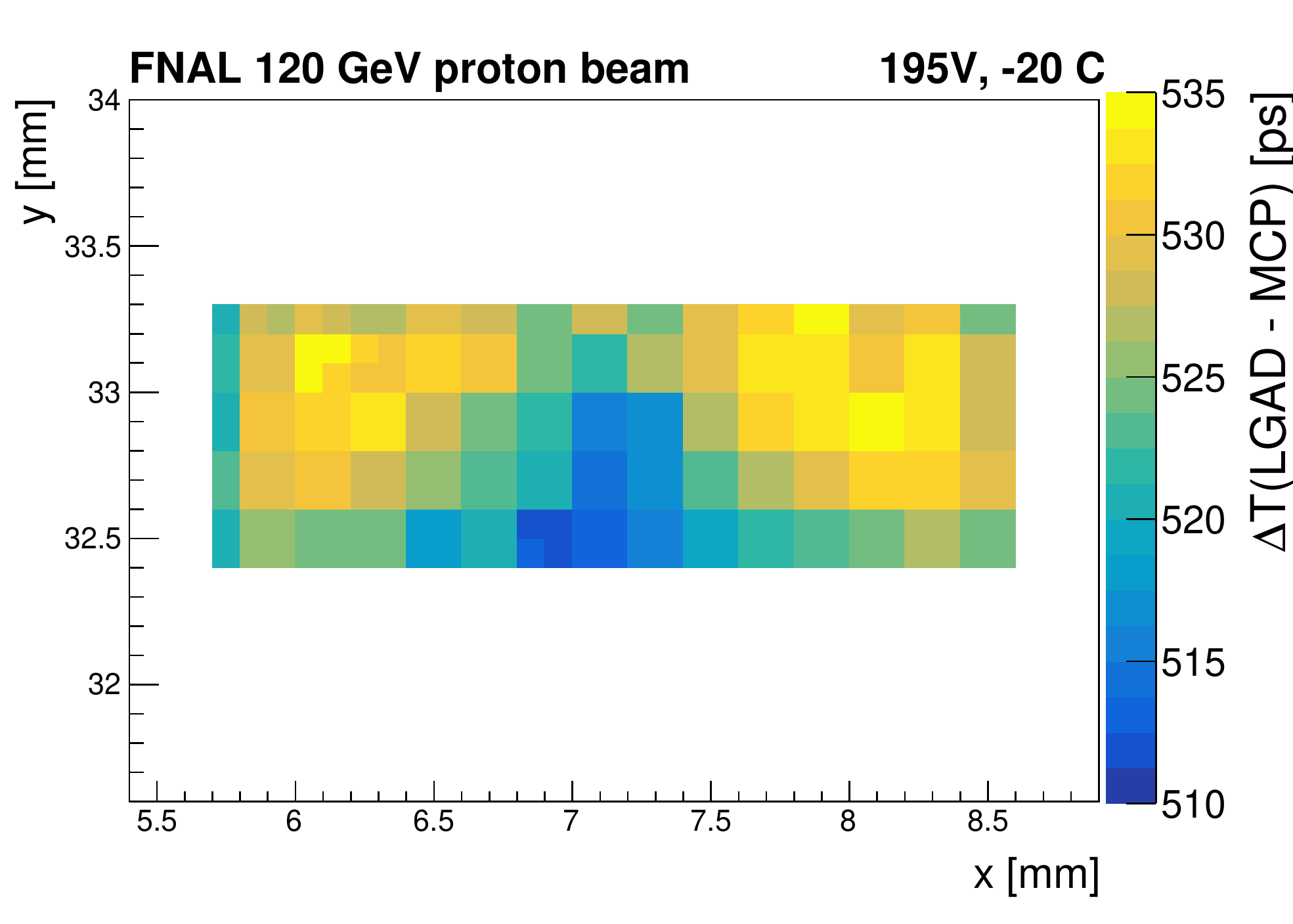} 

\caption{Average arrival time across a single pad after correcting for offsets in individual channels on the readout board, overlaying and averaging all pads together.} 
\label{fig:timedel_metalizedFeature} 
\end{figure}

The map of the measured signal arrival times is shown in Fig.~\ref{fig:timedel_4x4}. Due to differences in path lengths across the 16-channel board, there are different time offsets for each pad.
It was observed in a previous HPK sensor production~\cite{ApresyanLGAD} that a slight time offset exists within a single pad around the metalized bonding tab, shown in Fig.~\ref{fig:array_photo}. To measure the same feature in the HPK 3.1 sensor production, we
corrected the path-length offsets in each channel, and geometrically overlaid the signal arrival time measurements to visually enhance the feature around the bonding tab in Fig.~\ref{fig:timedel_metalizedFeature}. After this alignment, it is clear that the signals arising from protons passing underneath the metalized tab arrive approximately \SI{20}{\pico \s} earlier than other signals. The underlying cause for this feature is not entirely understood, but since the CMS and ATLAS timing detectors will use fully-metalized sensors, the time difference between metalized and non-metalized regions will not be relevant. Nonetheless, it is an interesting feature to monitor in future studies of sensors with limited surface metalization. 

The studies shown above demonstrate for the first time the operation of a large area, multi-pad LGAD sensor exposed to a particle beam. We observe a high degree of uniformity of signal amplitude and time resolution across the sensor surface and reliable operation over a period of several days.

\begin{figure}[htbp!] 
\centering
\includegraphics[width=0.49\textwidth]{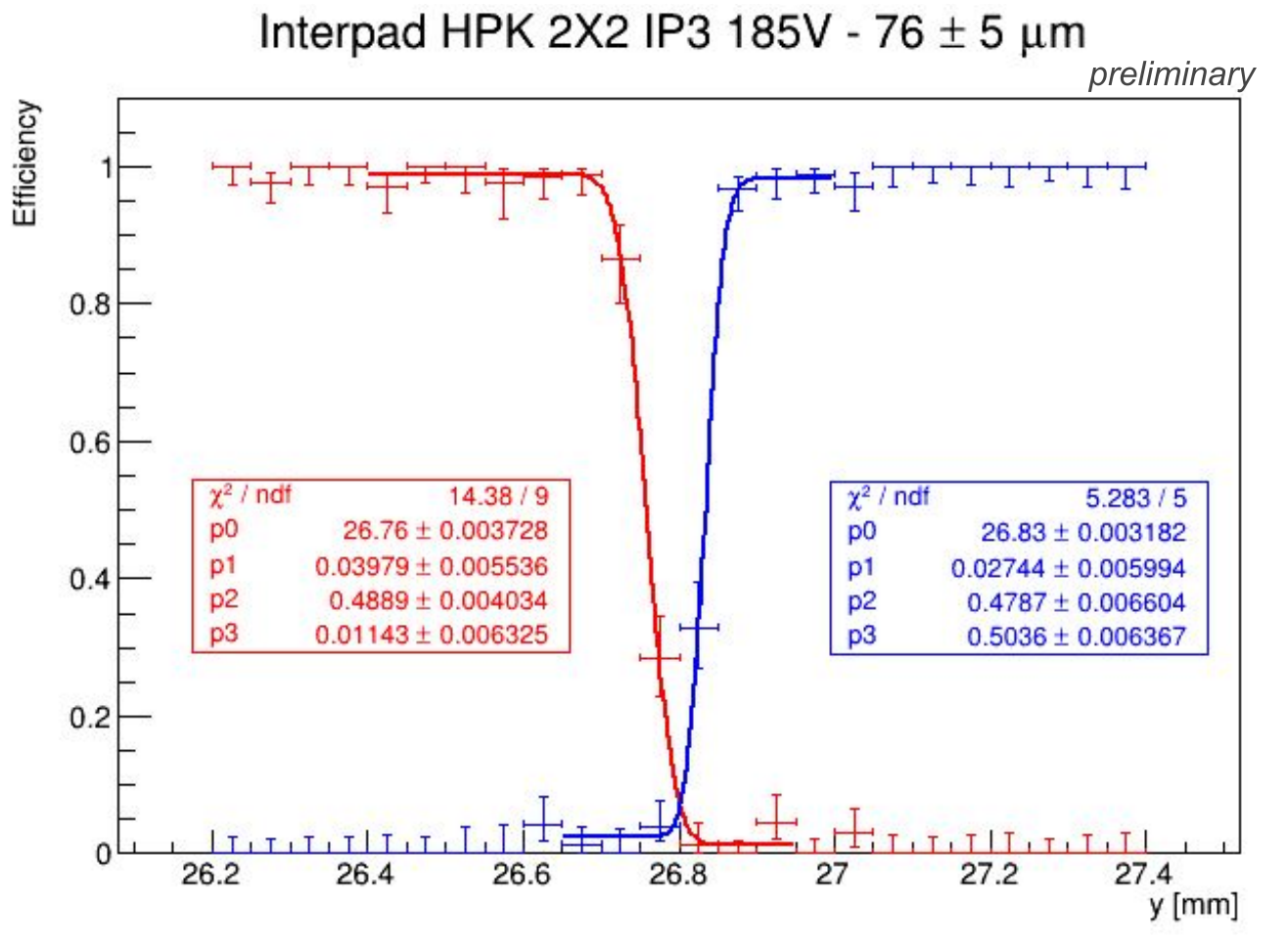}  
\includegraphics[width=0.49\textwidth]{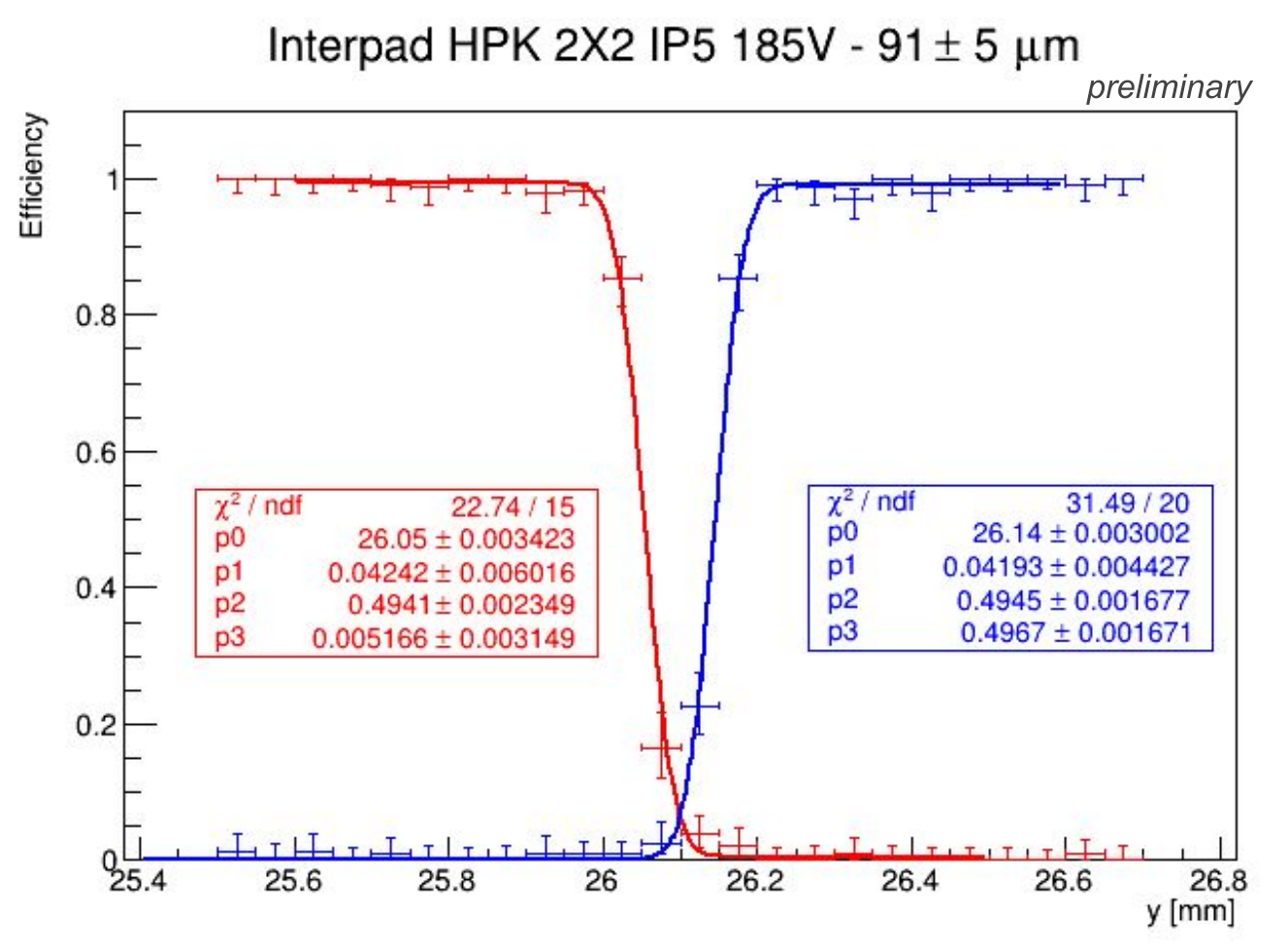}  
\includegraphics[width=0.49\textwidth]{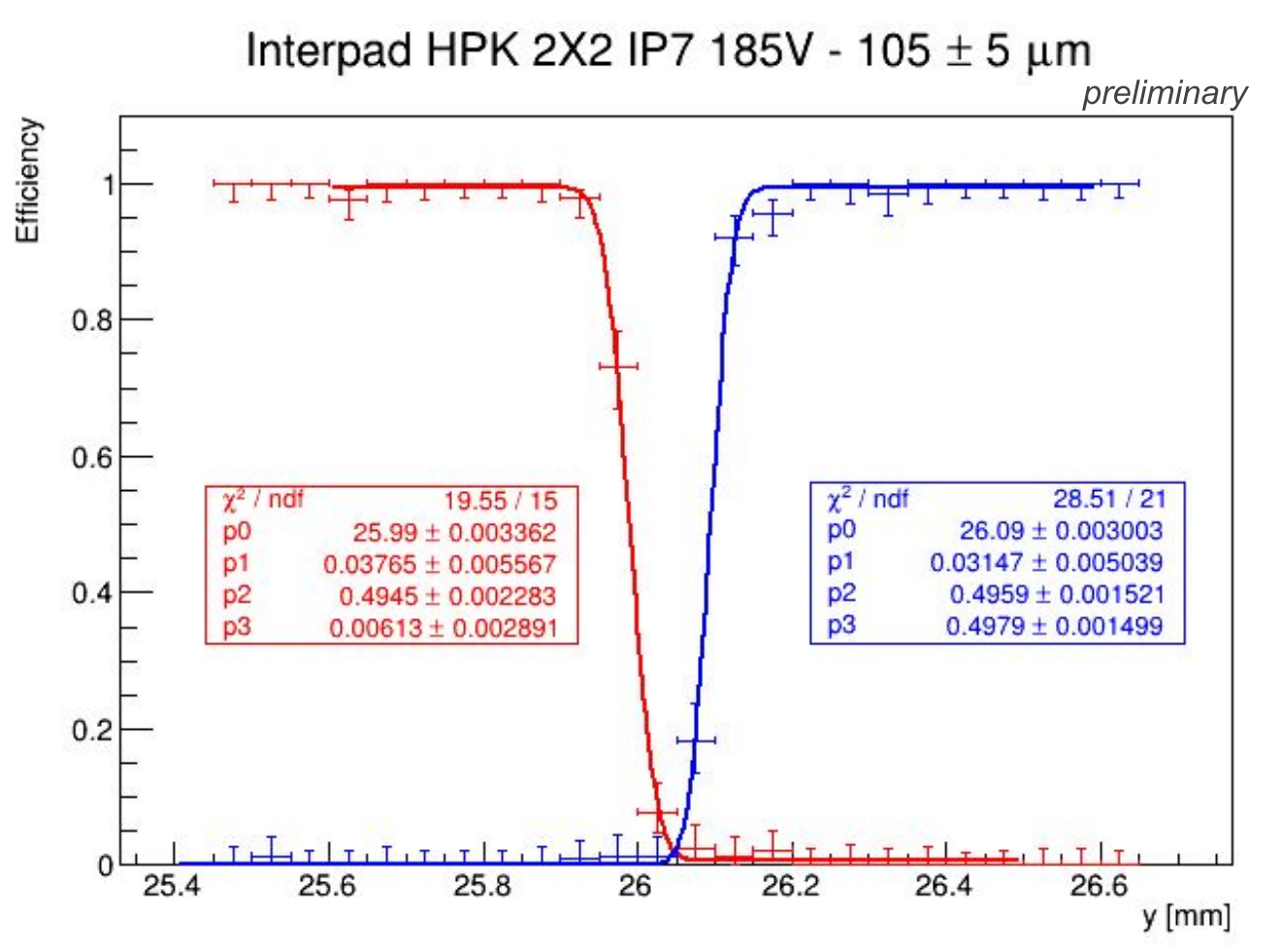}  
\caption{Measurement of the inter-pad gap in three different sensors. Top left: nominal gap \SI{30}{\micro \m}; measured 76 $\pm$ 5 \si{\micro \m}. Top right: nominal gap \SI{50}{\micro \m}; measured 91 $\pm$ 5 \si{\micro \m}. Bottom: nominal gap \SI{70}{\micro \m}; measured 105 $\pm$ 5 \si{\micro \m}.}
\label{fig:test_beam_IP} 
\end{figure}

Additionally, several similar sensors were studied with a variation in inter-pad gap width. Fig.~\ref{fig:test_beam_IP} shows 1D projections of the hit efficiency in the immediate vicinity of the inter-pad gaps. Each efficiency distribution is fit to a function which is a convolution of a step function representing the true efficiency, and a gaussian representing smearing from the tracker spatial resolution. 
The inter-pad gap is then defined as the distance between positions of 50\% efficiency on each fit function. 
The resulting inter-pad gaps are found to be consistent with the values obtained from laser TCT measurements performed previously \cite{BHARTHUAR2020164494}. This demonstrates that inter-pad gaps measured using the benchtop laser are consistent with the inter-pad gaps measured with particle signals as well. This is an important conclusion, as the benchtop measurements are generally easier to perform and have higher precision than can be achieved with the test beam.

\section{Beta source characterization campaign}
\label{sec:beta_campaign}

\begin{figure}[htbp!] 
\centering
\includegraphics[width=0.69\textwidth]{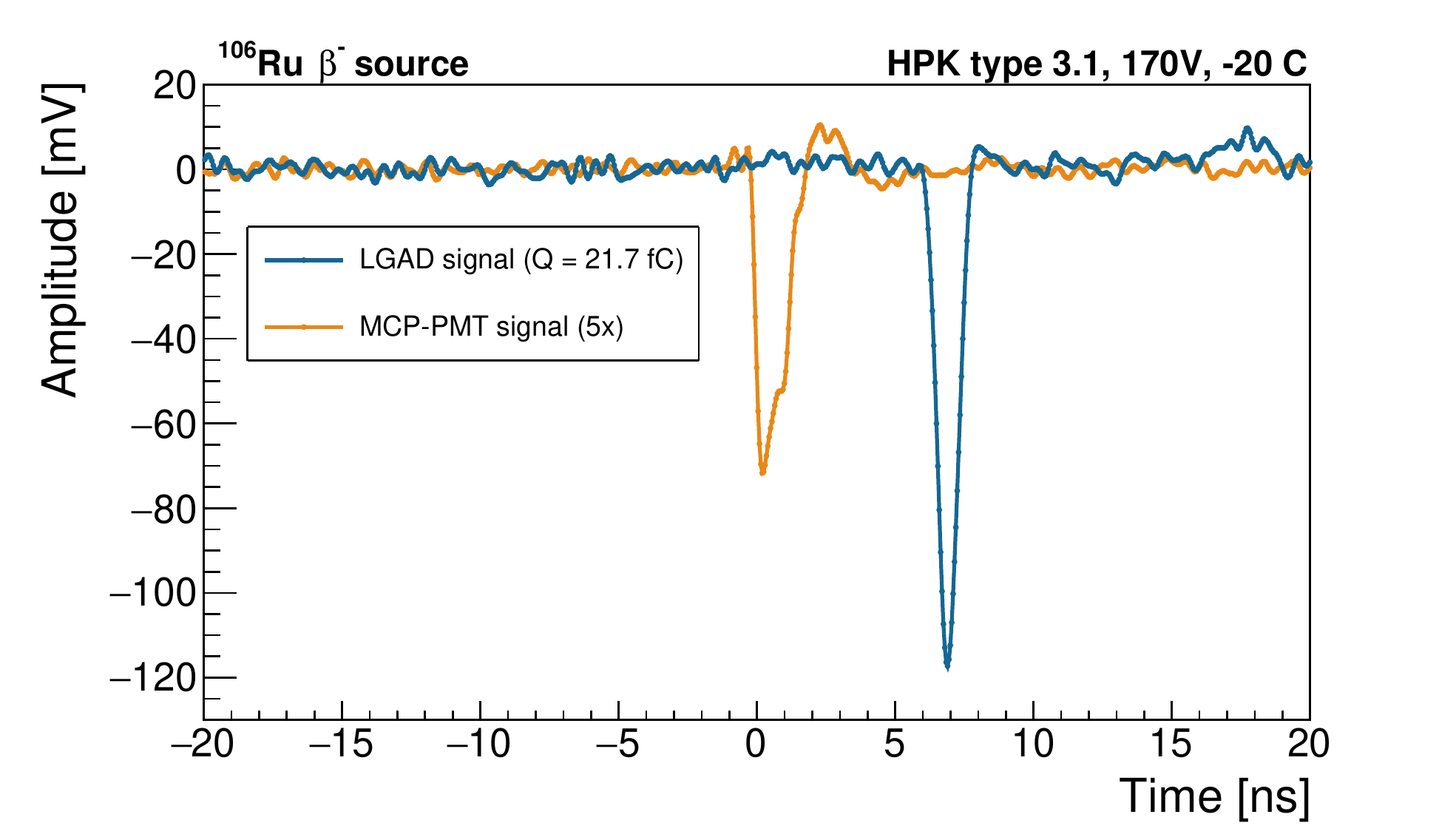}  
  
\caption{Waveforms from the LGAD and MCP-PMT in a representative event, sampled at 20 GS/s. The MCP-PMT waveform is scaled by a factor of 5 for better visibility. The LGAD waveform shown is from the sensor with \SI[product-units = power]{1 x 3}{\milli \meter} pads, at wafer position P2 with a nominal inter-pad gap of 50 microns (see Tab.~\ref{tab:sensors}). The collected charge observed in this event is \SI{21.7}{\femto \coulomb}, very close to the most probable value for this sensor (see distributions in Fig.~\ref{fig:tb_beta_hists}.) } 
\label{fig:event_display} 
\end{figure}

The beta source setup allows for much higher volume testing than is possible in the test beam. In this campaign, 22 HPK 3.1 sensors (as described in Section~\ref{sec:sensors}) were characterized with the beta source. One pad on each sensor was read out using the UCSC board.  Each sensor was subjected to a bias voltage scan in increments of \SI{5}{\volt} up to \SIrange{5}{10}{\volt} less than the breakdown voltage.  At each bias voltage point, a few thousand events are acquired, and the entire process takes approximately one day to complete. A representative event collected using the beta source is shown in Fig.~\ref{fig:event_display}, including both the LGAD and MCP-PMT waveforms.

\subsection{Comparison of beta particle and proton response}
 \label{ssec:tb_beta_validation}

\begin{figure}[htbp!] 
\centering
\includegraphics[width=0.49\textwidth]{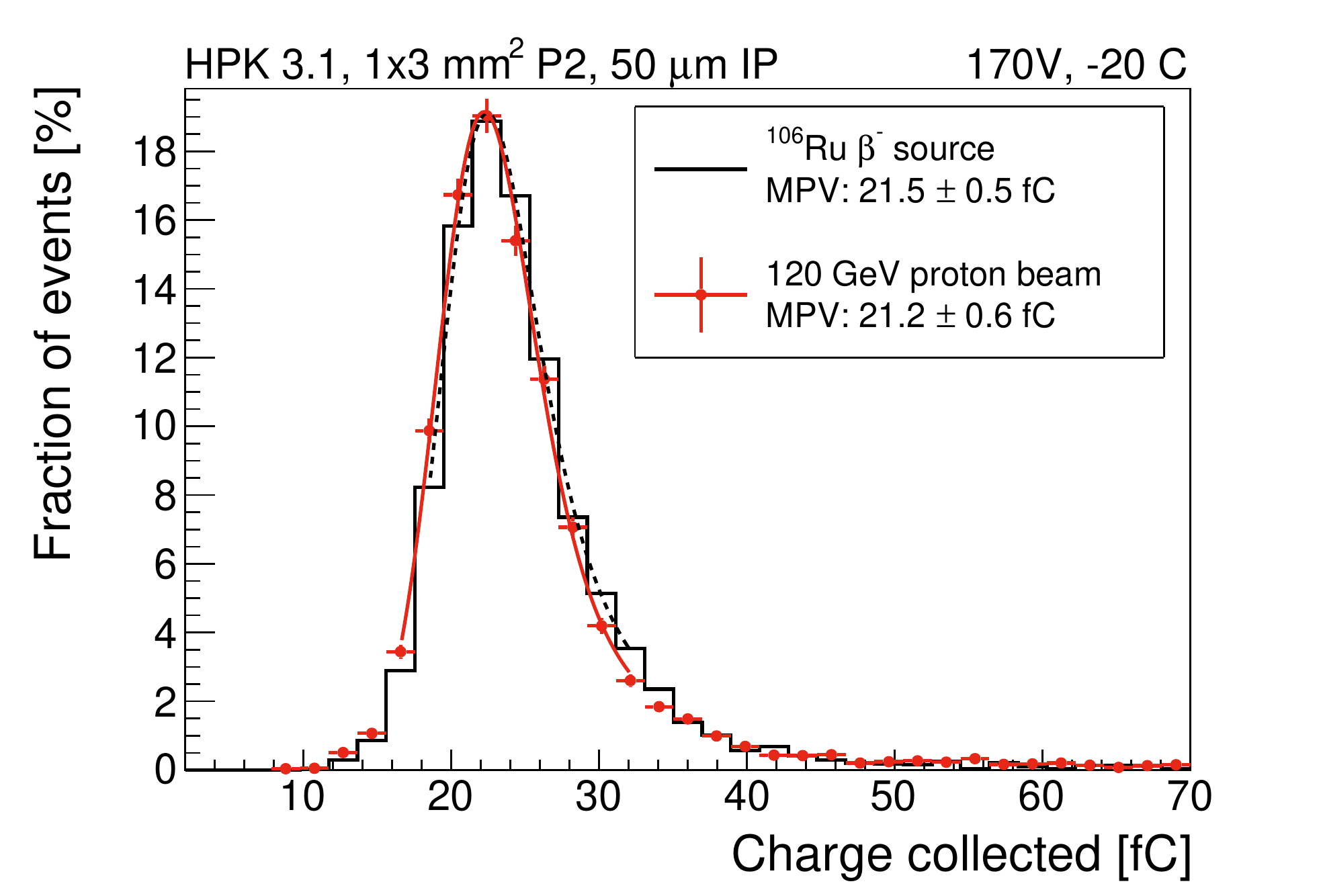}  
\includegraphics[width=0.49\textwidth]{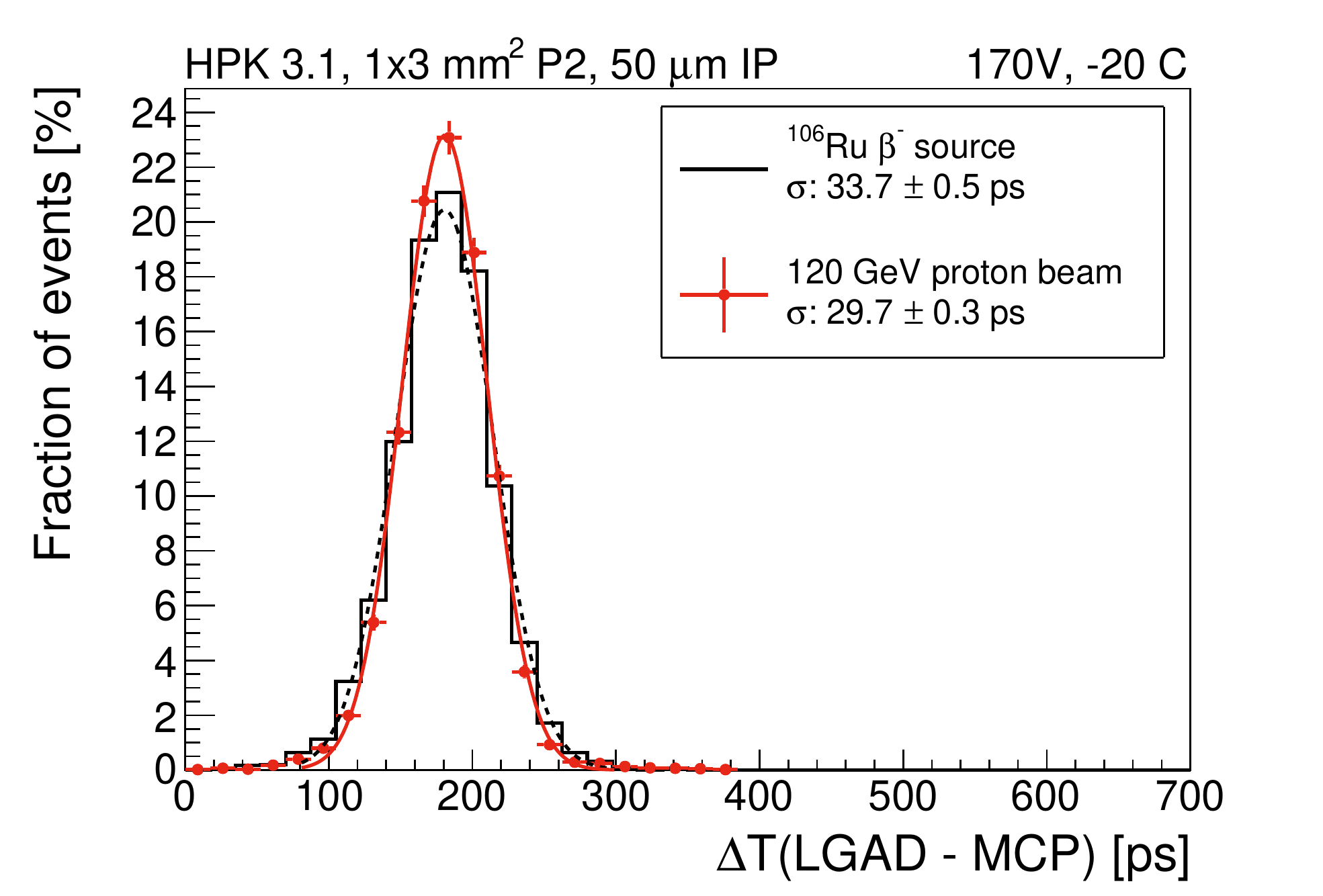}   
\caption{Distributions of collected charge (left) and arrival time difference (right) observed from a single sensor exposed to the test beam and the beta source. This sensor has a pad size of \SI[product-units = power]{1 x 3}{\milli \meter}, is from wafer position P2, and has a nominal inter-pad gap of 50 microns. It was operated at a bias voltage of \SI{170}{\volt} and a temperature of -20 C. The legends indicate the most probable values of the collected charge, and the width of the $\Delta \text{T}$ distributions. The width includes contributions from the LGAD resolution as well as the MCP-PMT reference.} 
\label{fig:tb_beta_hists} 
\end{figure}

\begin{figure}[htbp!] 
\centering
\includegraphics[width=0.49\textwidth]{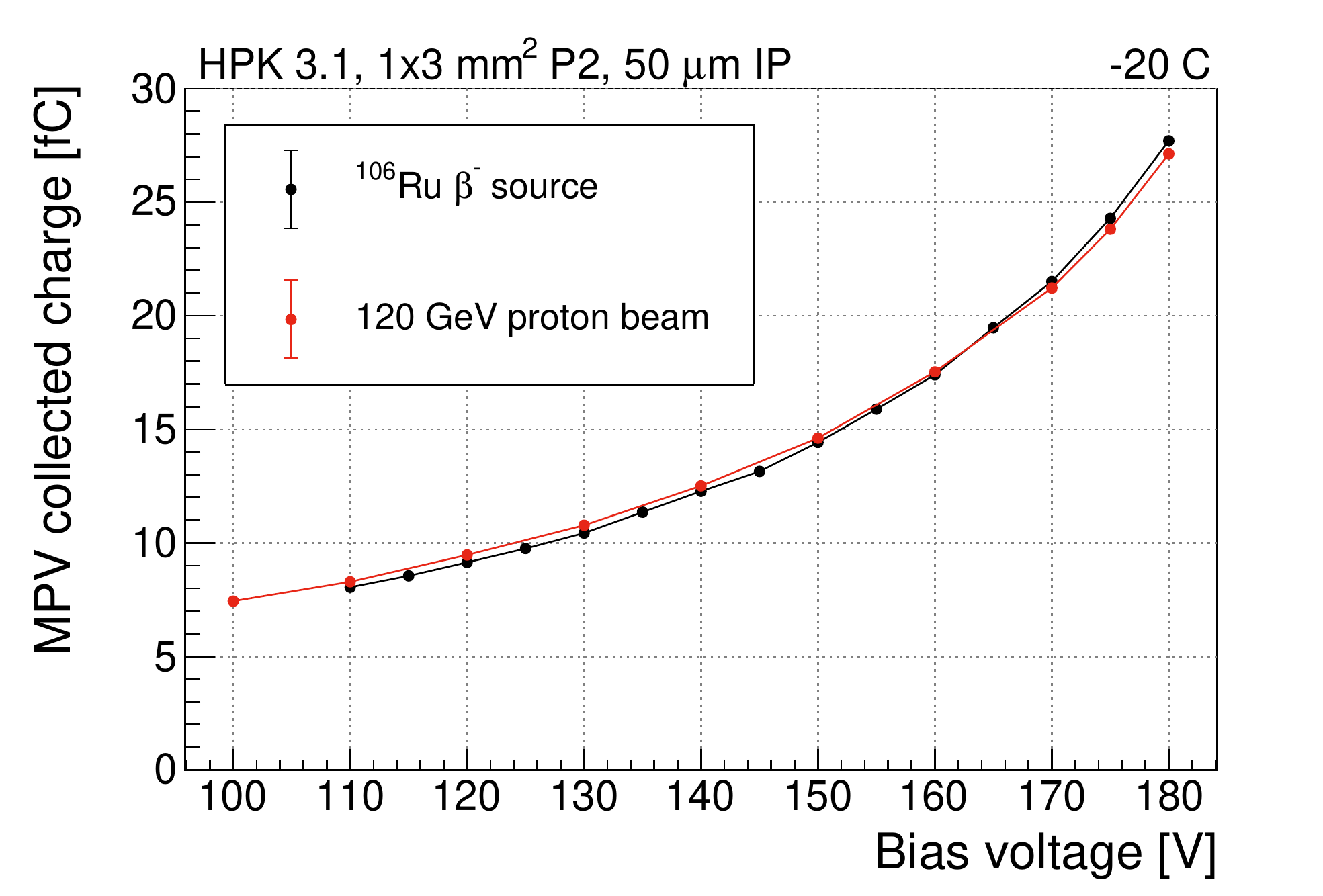}  
\includegraphics[width=0.49\textwidth]{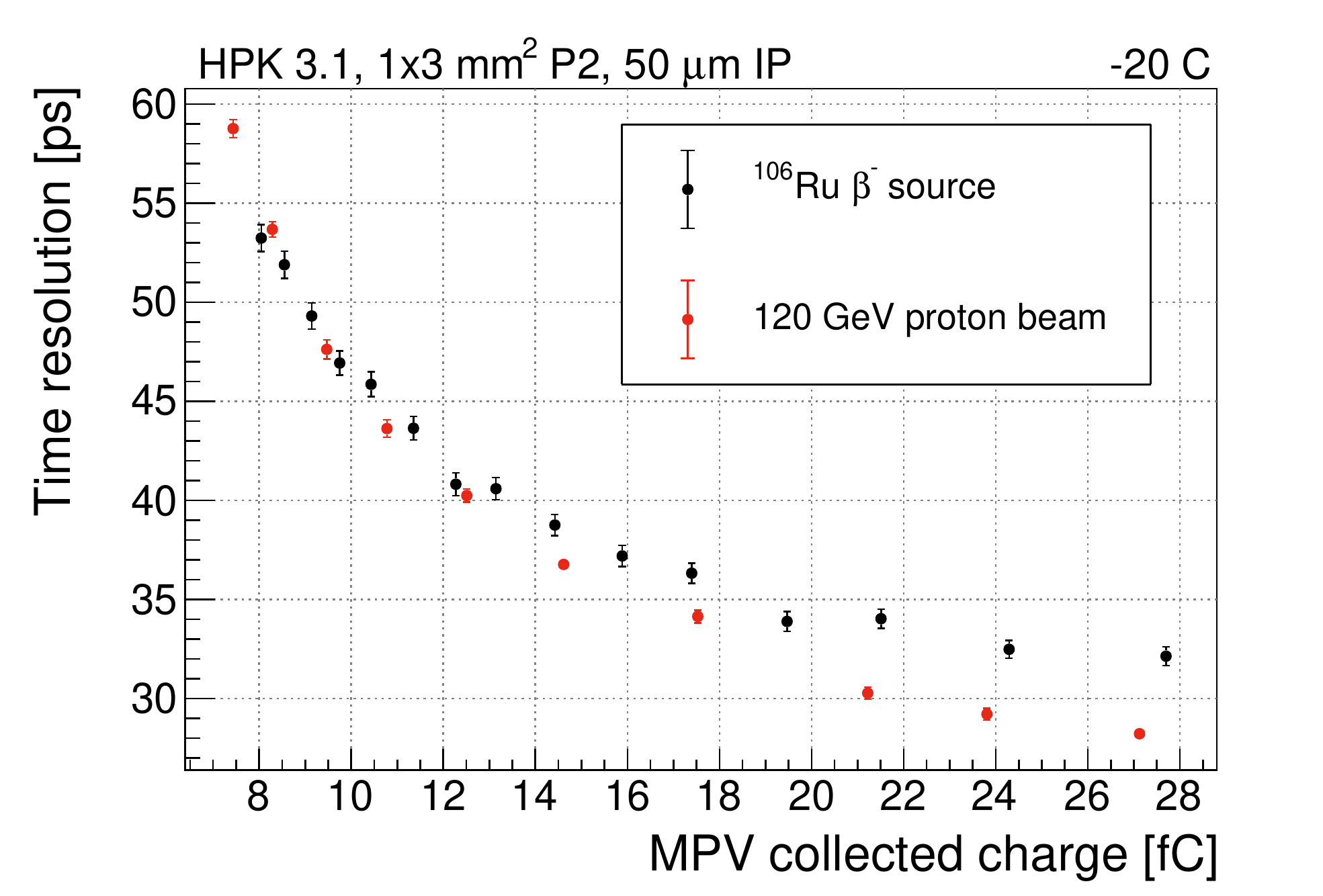}   
\caption{Comparison of the collected charge (left) and time resolution (right) observed using the test beam and the beta source for a sequence of measurements with varying bias voltage. The measurements with a bias voltage of \SI{170}{\volt} correspond to the distributions shown in Fig.~\ref{fig:tb_beta_hists}. The time resolutions shown includes both the LGAD and MCP-PMT contributions. The small difference observed at very high charge is used to estimate the additional resolution contribution present in the beta source setup.} 
\label{fig:tb_beta_comparison} 
\end{figure}

For the applications in CMS and ATLAS, the LGAD response to minimum ionizing particles (MIP) is the primary concern. Beta particles with 1 MeV energy are minimally ionizing for passage through small amounts of material~\cite{PDG}. As a first step to commission the beta source setup, we assess how well the signals arising from beta particles correspond with those from MIPs in the test beam. For this demonstration, we studied one sensor in an identical configuration using both the beta source and the \SI{120}{\GeV} proton beam. The distributions of charge and arrival time for a bias voltage of \SI{170}{\volt} are shown in Fig.~\ref{fig:tb_beta_hists}, and their evolution as a function of bias voltage is shown in Fig.~\ref{fig:tb_beta_comparison}.
The charge depositions and signal shapes originating from the two signal types are in remarkably good agreement, which indicates that the selected beta events do represent a population of minimum ionizing particles. As previously described, the requirement that the beta particles pass in a straight path through the pinhole and reach the MCP-PMT reference detector ensures that particles experiencing larger energy loss or a hard scatter are excluded. 

Although the LGAD signal properties match very well between the beta source and proton datasets, it can be seen in Figs.~\ref{fig:tb_beta_hists} and~\ref{fig:tb_beta_comparison} that the resolutions measured from the beta source dataset is consistently larger by about 15~\si{\pico \s} in quadrature. This degradation is due to two factors: the reduced response of the MCP-PMT reference to beta particles, and variations in path length (time of flight) that are present in the beta setup but not the test beam. A 15~\si{\pico \s} contribution is subtracted in quadrature from all subsequent resolution measurements in order to isolate the resolution of the LGAD detector. 


 \subsection{Relative calibration of readout boards}
 \label{ssec:board_cal}

\begin{figure}[htbp!] 
\centering

\includegraphics[width=0.49\textwidth]{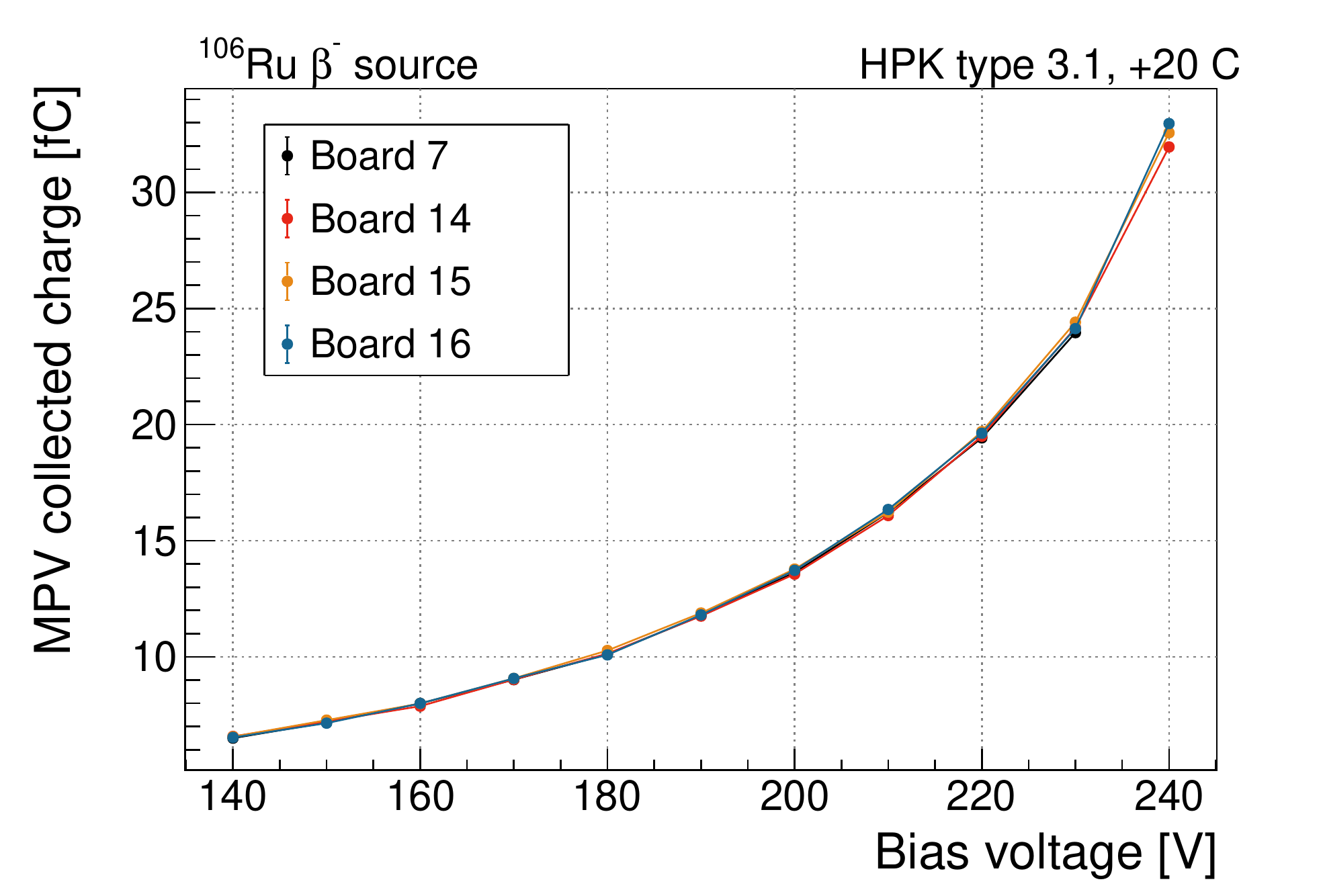}  
\includegraphics[width=0.49\textwidth]{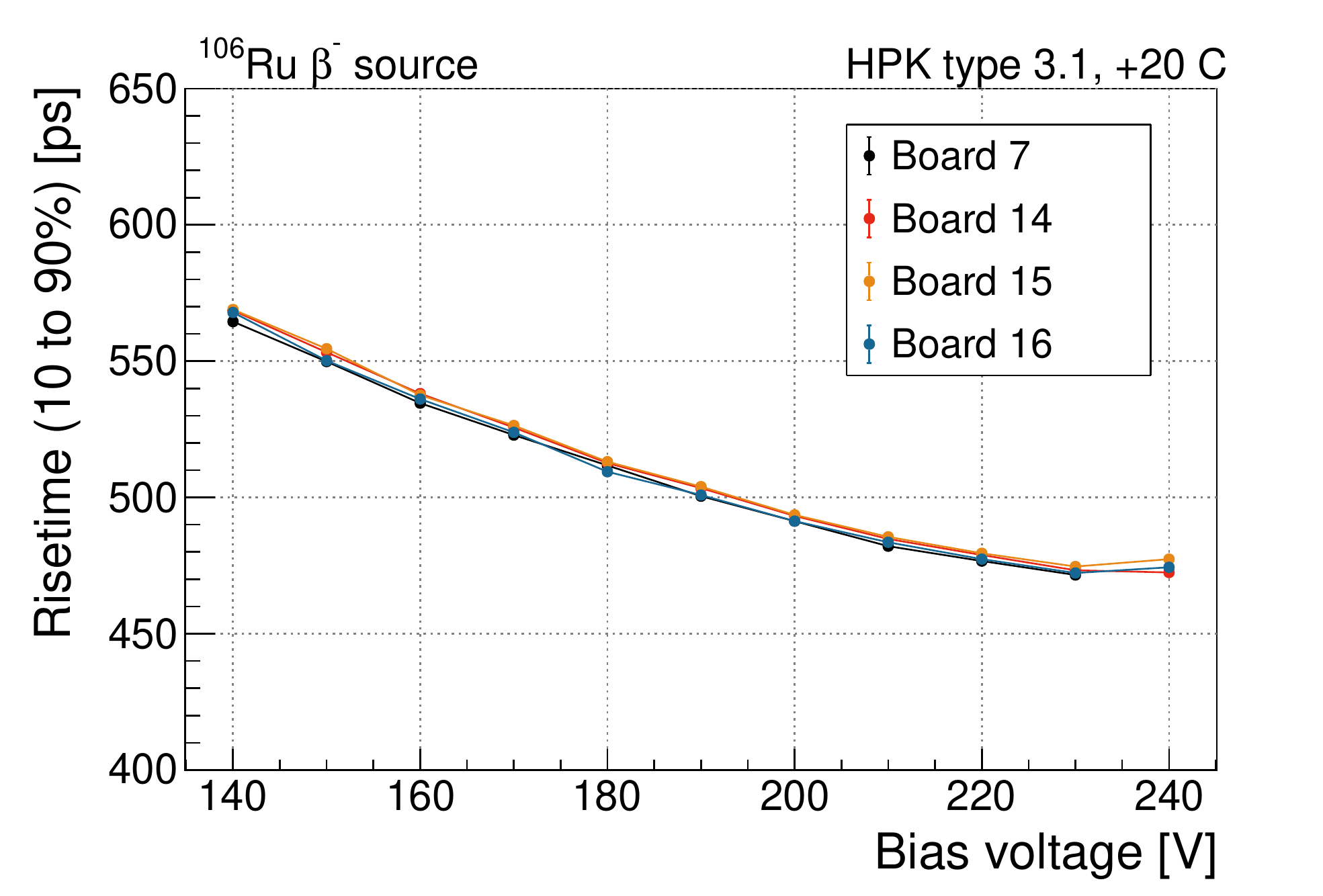}  
\caption{Comparison of the four UCSC readout boards used for the beta source characterization campaign. All four show good uniformity in charge (left) and risetime (right) for a single sensor rotated through all four boards.} 
\label{fig:board_cal} 
\end{figure}

To efficiently characterize a large population of sensors, we employed multiple copies of the UCSC readout board to enable mounting and wirebonding sensors in batches. It is then important to establish that all readout boards give uniform results. Before the campaign began, a single calibration LGAD was cycled through all candidate readout boards to gauge the board uniformity. In fact, all four of the UCSC readout boards used in this campaign reproduced essentially identical results, as can be seen in Fig.~\ref{fig:board_cal}. The remarkable reproducibility of the results demonstrates the excellent environmental control in the beta setup and builds confidence that even small differences observed between sensors are meaningful.

\subsection{Beta source results}
\label{ssec:beta_results}
\begin{figure}[htp] 
\centering
\includegraphics[width=0.49\textwidth]{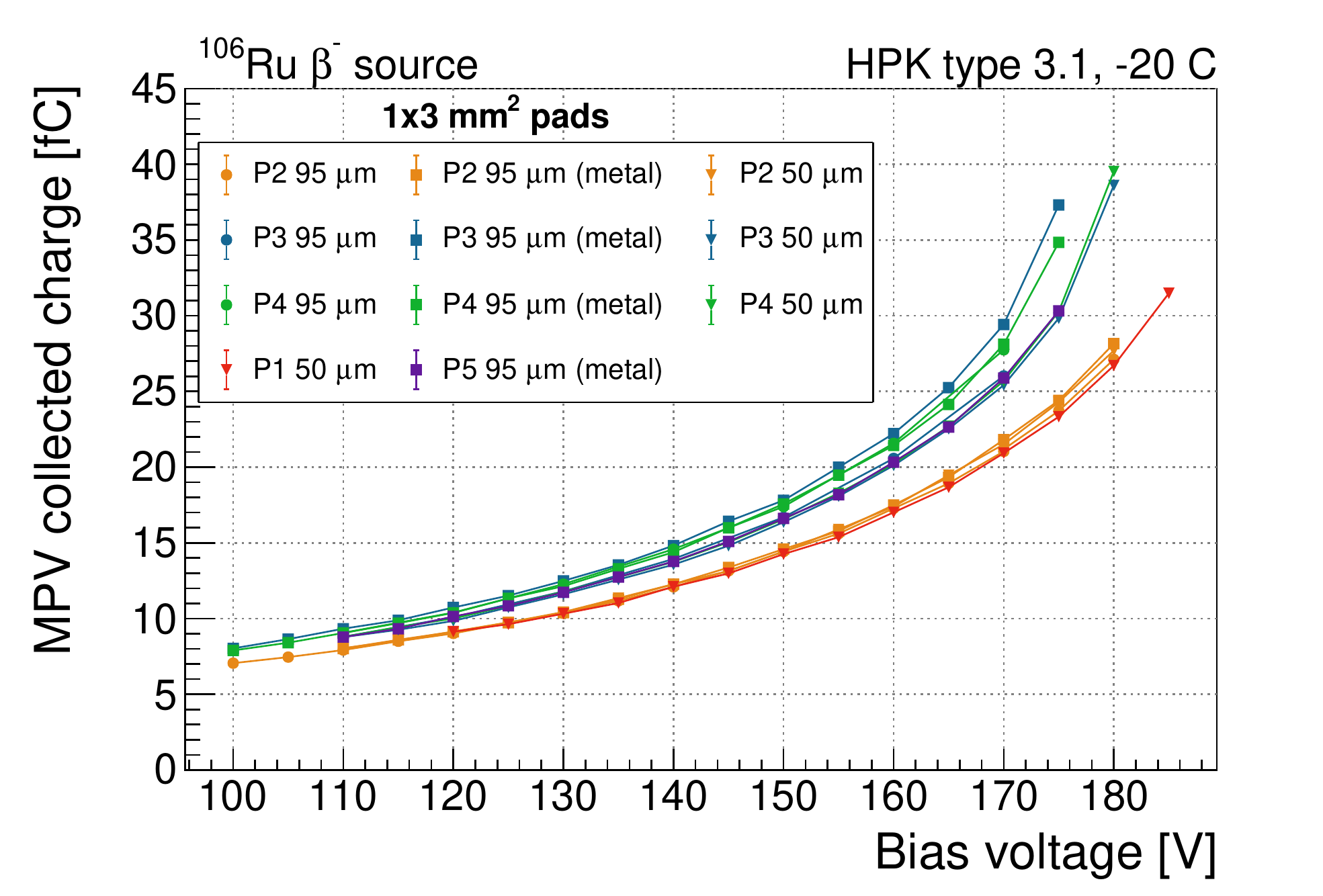}  
\includegraphics[width=0.43\textwidth]{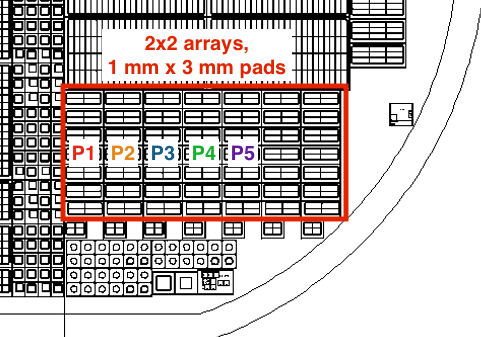} \caption{Collected charge as a function of bias for the sensors with \SI[product-units = power]{1 x 3}{\milli \meter} pads (left). The legend indicates the nominal inter-pad gap values, whether the surface is fully metalized, the position on the wafer (P1-P5). The right panel shows a portion of the wafer layout from Fig.~\ref{fig:wafer_layout} that indicates the locations of columns P1-P5.  The pads near the center of the wafer (P1-P2) need an additional \SI{10}{\volt} to deliver the same charge as the sensors from the edge of the wafer (P3-P5).} 
\label{fig:beta_charge_vs_bias} 
\end{figure}

\begin{figure}[htp] 
\centering
\includegraphics[width=0.49\textwidth]{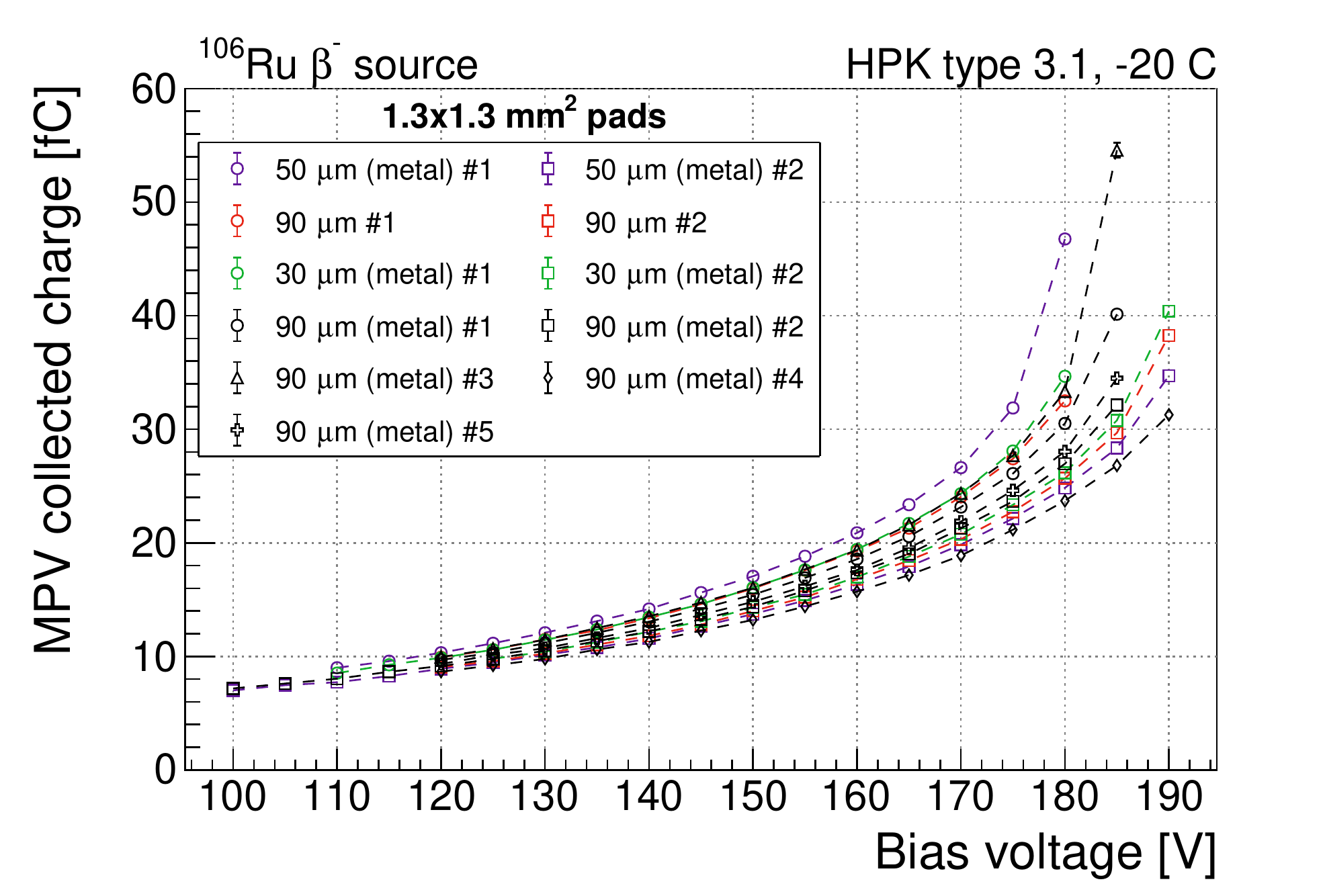}  
\caption{Collected charge as a function of bias for the sensors with \SI[product-units = power]{1.3 x 1.3}{\milli \meter} pads (right). The legend indicates the nominal inter-pad gap values, whether the surface is fully metalized.} 
\label{fig:beta_charge_vs_bias2} 
\end{figure}

\begin{figure}[htp] 
\centering
\includegraphics[width=0.49\textwidth]{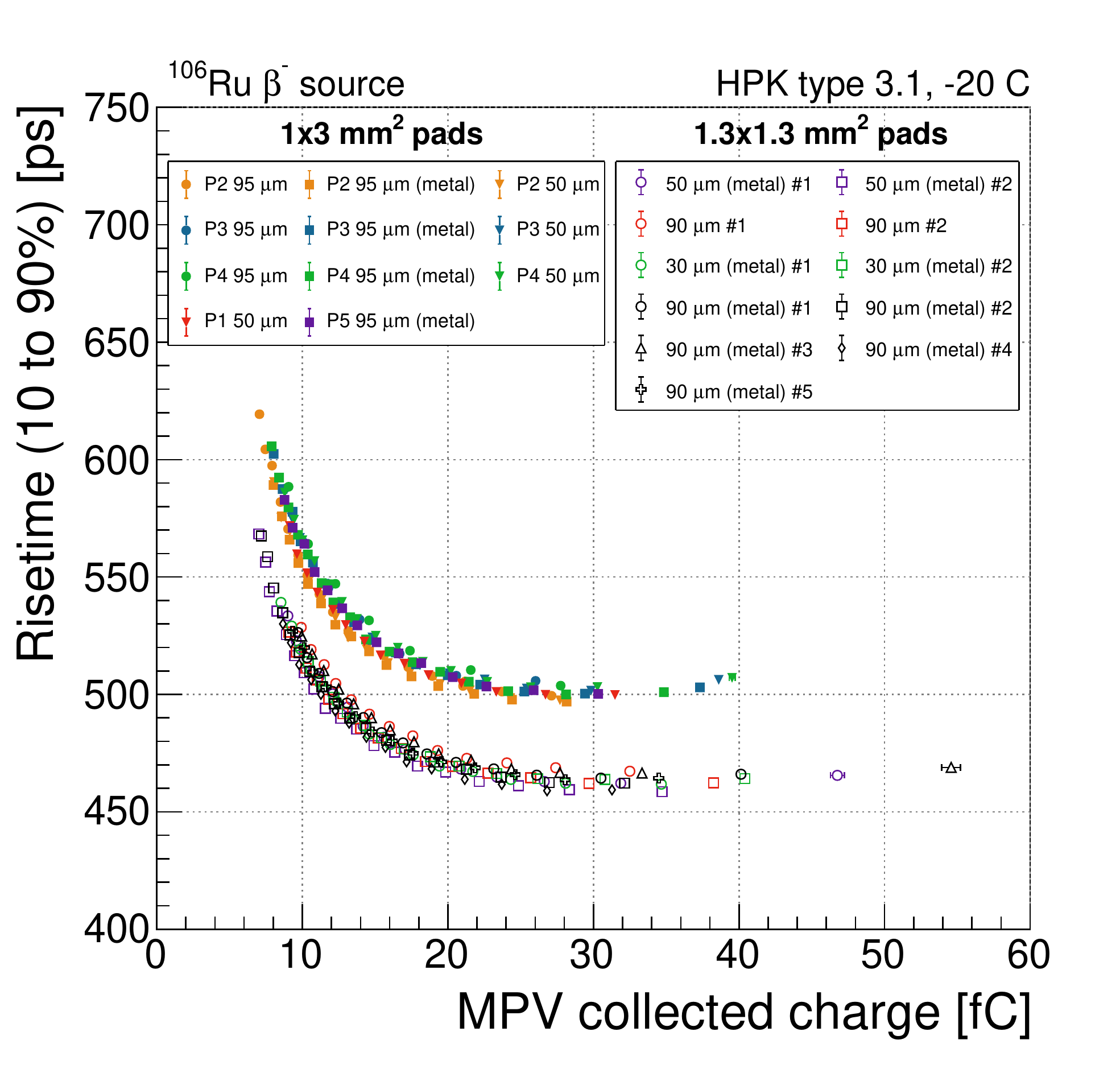}  
\includegraphics[width=0.49\textwidth]{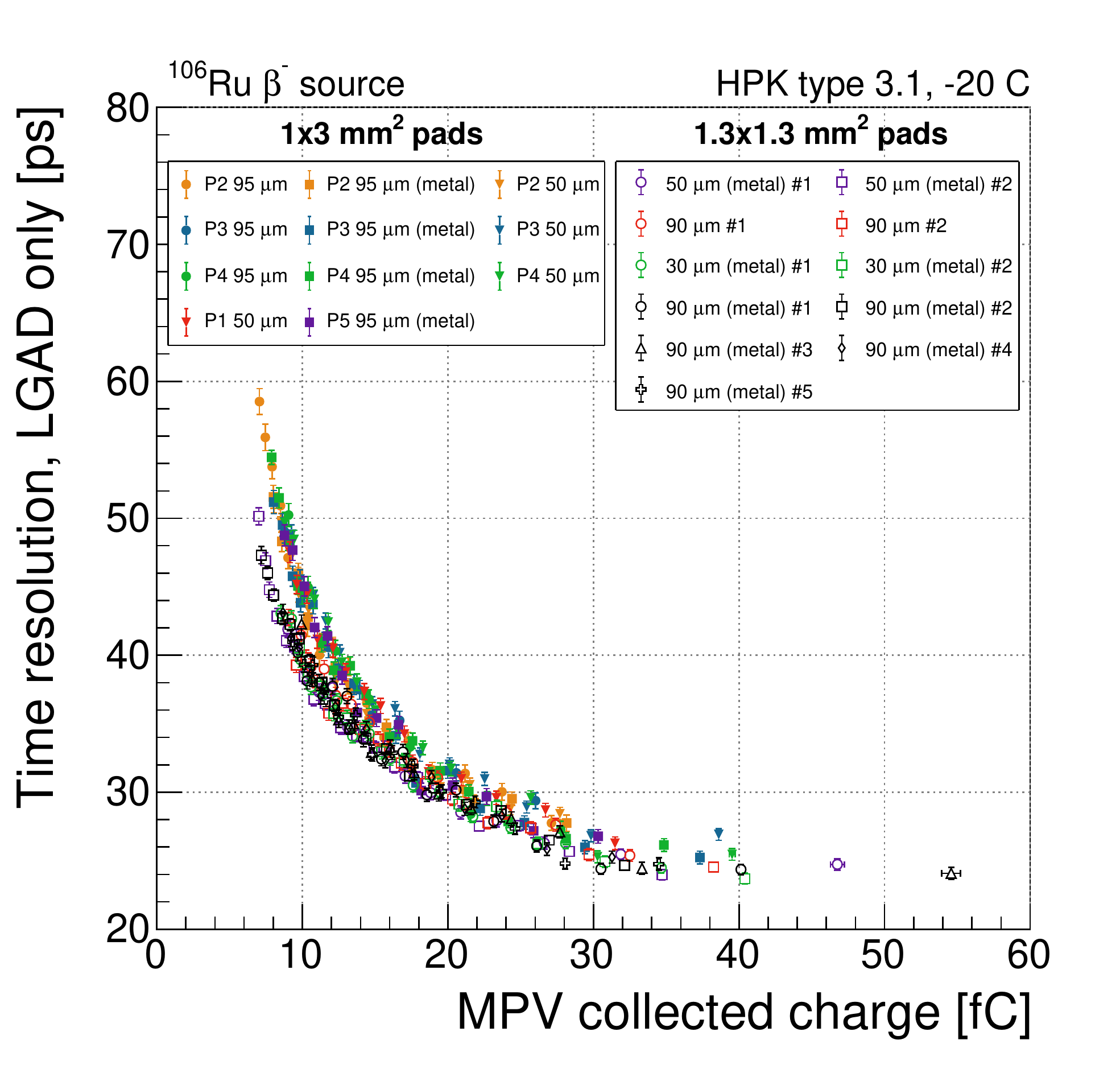}  
\caption{Risetime and time resolution for the 22 sensors as a function of the collected charge at each bias voltage. The \SI[product-units = power]{1 x 3}{\milli \meter} pads have a slightly slower risetime due to the larger capacitance, but similar asymptotic time resolution at high values of charge. The time resolution values shown have been corrected to remove the MCP-PMT contribution of \SI{15}{\pico \second}.}
\label{fig:beta_time_vs_charge} 
\end{figure}

The results of the beta measurement campaign are shown in Figures~\ref{fig:beta_charge_vs_bias} to ~\ref{fig:beta_CV_measurements}. Figures~\ref{fig:beta_charge_vs_bias} and ~\ref{fig:beta_charge_vs_bias2} show the charge collection as a function of bias voltage for the sensors of pad size \SI[product-units = power]{1 x 3}{\milli \meter} and \SI[product-units = power]{1.3 x 1.3}{\milli \meter}. Variation in the gain layer concentration between sensors results in a translation of these curves along the bias voltage axis, with the key figure of merit being the bias voltage to reach a certain charge threshold. Among the \SI[product-units = power]{1 x 3}{\milli \meter} sensors, there is a correlation with the turn-on of the charge curve and the position of the sensor on the wafer. Sensors near the center of the wafer (P1-P2) require a bias voltage approximately \SI{10}{\volt} larger to reach the same gain as sensors towards the edge of the wafer (P3-P5). The wafer positions of the \SI[product-units = power]{1.3 x 1.3}{\milli \meter} sensors were not preserved, but a similar scale of variation in bias voltage of 10--15\si{\volt} is observed to reach a given gain value. These differences represent slight variation in the concentration of the gain implant.

Figure~\ref{fig:beta_time_vs_charge} shows the risetime and time resolution for each LGAD as a function of the collected charge at each bias point. There are two groups observed in the risetime distribution, corresponding to pads with \SI[product-units = power]{1.3 x 1.3}{\milli \meter}  and \SI[product-units = power]{1 x 3}{\milli \meter} areas. The \SI[product-units = power]{1.3 x 1.3}{\milli \meter} pads reach \SI{460}{\pico \s} and \SI[product-units = power]{1 x 3}{\milli \meter}  pads reach \SI{500}{\pico \s} risetimes. This offset is expected due to the different capacitances introduced by each pad size. The faster risetime yields improved time resolution at low charge, but when operated at high gain, both pad sizes converge to an asymptotic time resolution of 25--30\si{\pico \s}. Within each of the two sensor populations, the relationship between collected charge and either risetime or time resolution is common to all sensors. The small differences in operating voltage, visible in Figs.~\ref{fig:beta_charge_vs_bias} and ~\ref{fig:beta_charge_vs_bias2}, have limited impact on timing performance at any given charge. There is a small effect from the difference in operating voltage: sensors with a larger operating voltage have risetime that is about \SI{10}{\pico \s} faster than sensors with a lower operating voltage. This can be seen by comparing the P2 (orange) sensors against the P4 (green) sensors in Fig.~\ref{fig:beta_charge_vs_bias} and ~\ref{fig:beta_time_vs_charge} (left). This stems from the fact that electron drift velocity is not yet saturated until the highest bias voltages, and so the sensors with higher operating voltages are closer to velocity saturation at a specific gain.

As can be seen in Fig.~\ref{fig:beta_time_vs_charge}, a charge collection of 20 \si{\femto\coulomb} is sufficient to provide \SI{30}{\pico\second} or better time resolution. We define the voltage where 20 \si{\femto\coulomb} charge collection is achieved to be the minimum operating voltage of each sensor. Furthermore, from Fig.~\ref{fig:beta_charge_vs_bias}, we see that the sensor can still be operated with a bias up to about \SI{20}{\volt} above the minimum operating voltage, before entering breakdown region. Therefore, each sensor has a roughly \SI{20}{\volt} bias voltage range for operation with better than \SI{30}{\pico\second} time resolution.

Sensors were studied with variation in inter-pad gap size, and both with and without surface metalization. We observe that neither of these properties have a significant effect on any aspect of the performance, as can be seen in Fig.~\ref{fig:beta_charge_vs_bias} and~\ref{fig:beta_time_vs_charge}. 


\subsection{Correspondence with probe CV measurements}
\label{ssec:probeCV}
A key challenge for the operation of a large scale timing detector is ensuring all pads on each sensor can reach the gain needed to achieve the required time resolution with all pads on each sensor constrained to operate at the same bias voltage. Due to the long duration and complexity of the measurement, it is not possible to characterize each sensor fully using the beta source telescope during the construction phase of the detector. However, probe station measurements for a sample of pads on each sensor are a practical alternative for sensor quality characterization. Therefore establishing the relationship between charge collection and the probe station measurements allows for efficient LGAD sensor quality control and characterization of uniformity. With the goal of establishing that relationship, the 22 sensors included in this campaign were characterized using a probe station in the Torino UFSD lab~\cite{Sola:2667026}.

\begin{figure}[htbp!] 
\centering
\includegraphics[width=0.49\textwidth]{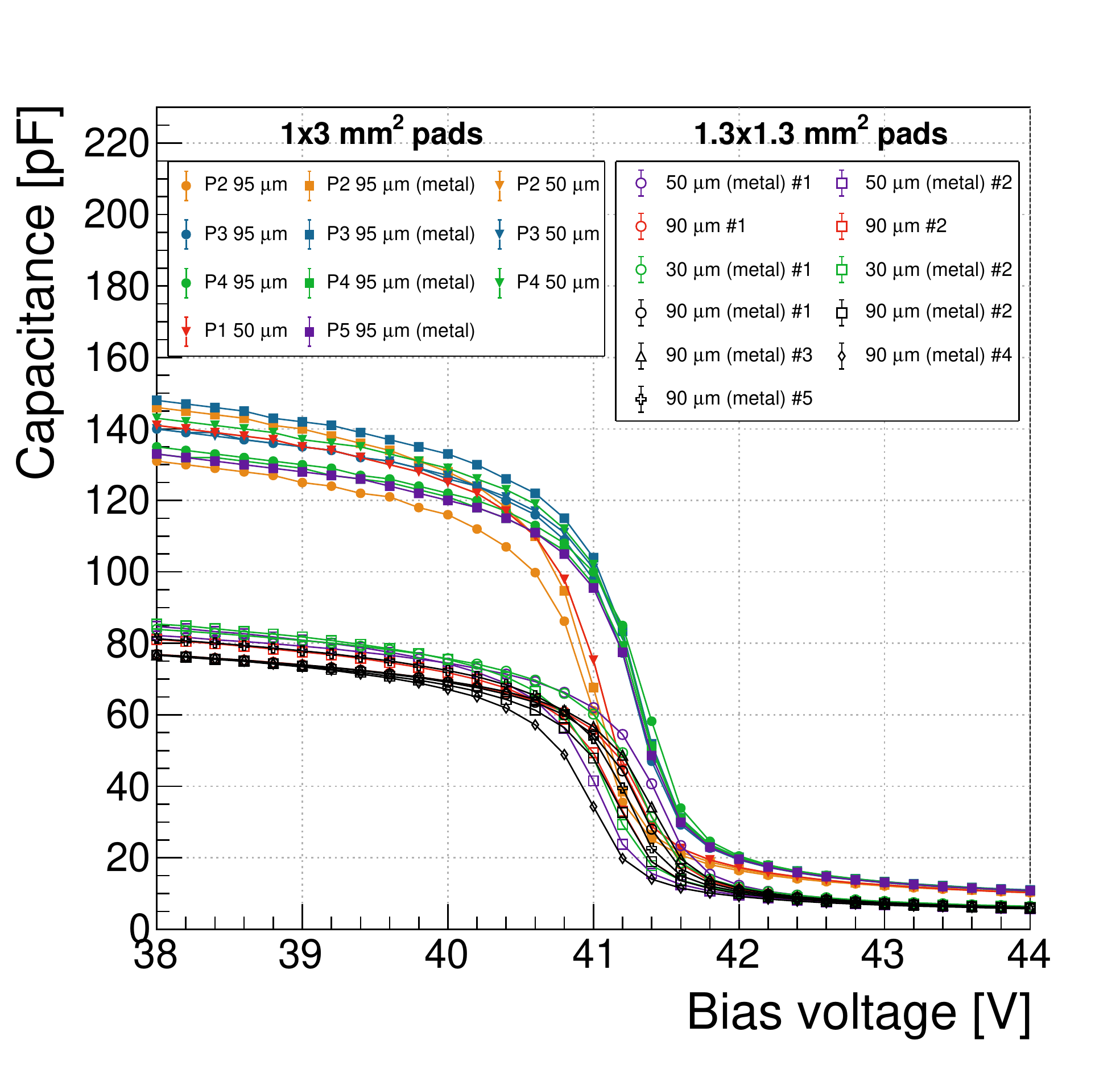}  
\includegraphics[width=0.49\textwidth]{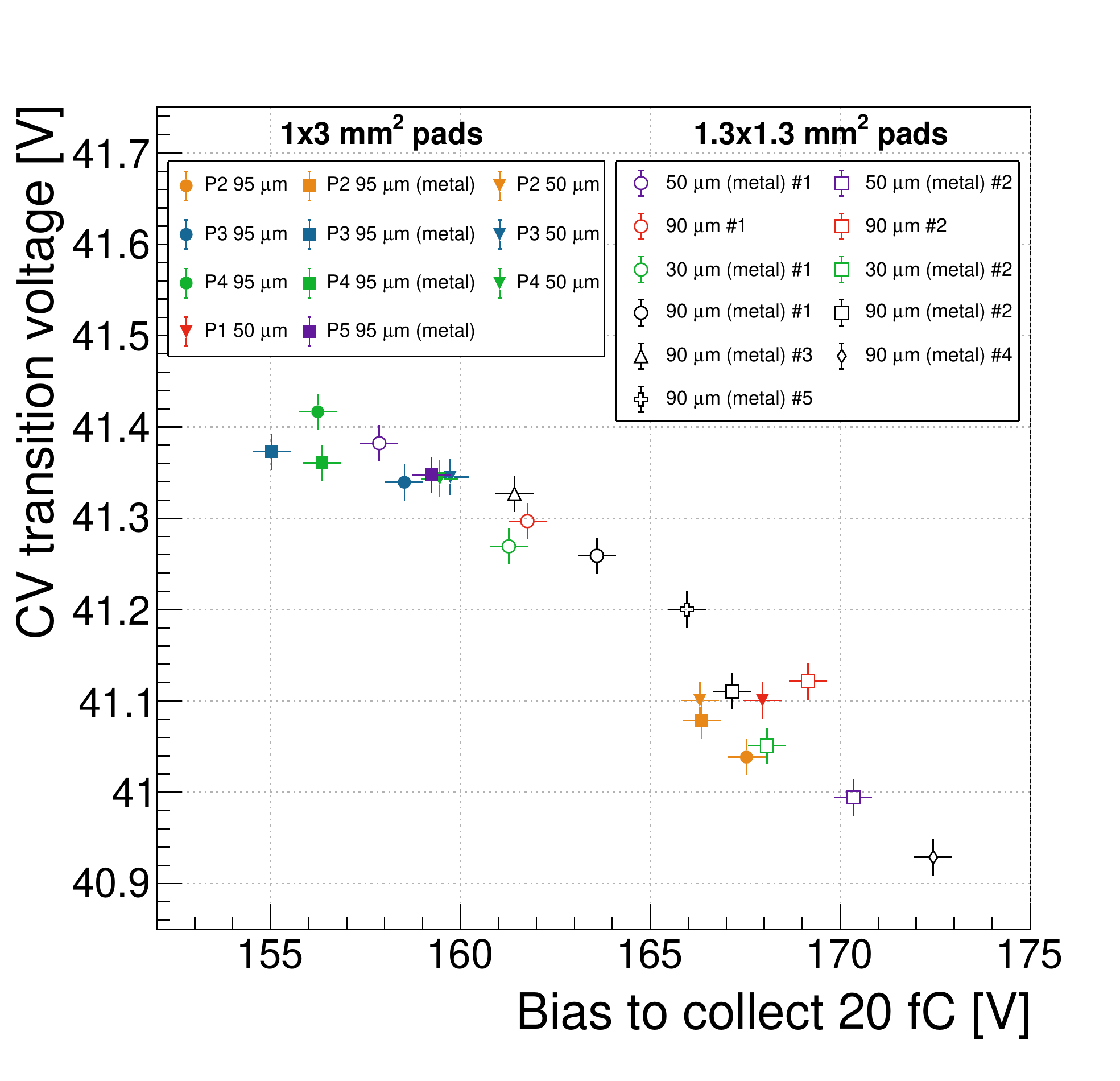}  
\caption{Capacitance-voltage curve for the 22 sensors, measured with probe station (left). Correlation of the CV transition voltage of each sensor with the bias voltage to collect \SI{20}{\femto \coulomb} (right).} 
\label{fig:beta_CV_measurements} 
\end{figure}

Probe station CV measurements can be used to determine the depletion voltage of the gain layer, which indicates the concentration of the gain layer dopant. The CV curves for all 22 sensors are shown in Fig.~\ref{fig:beta_CV_measurements} (left). Sensors which require a larger voltage to deplete the gain layer should have a larger gain at a given bias voltage. To quantify the variation in the depletion voltage of the gain layer, we define a CV transition voltage where the capacitance crosses a particular threshold: \SI{75}{\pico \farad} (\SI{52}{\pico \farad}) for \SI[product-units = power]{1 x 3}{\milli \meter} (\SI[product-units = power]{1.3 x 1.3}{\milli \meter}) pad sensors. These capacitance thresholds correspond to roughly the midpoint of the steeply falling portions of each curve.

We study the relationship between the CV transition voltage and the operating voltage of each sensor. As was observed in Sec.~\ref{ssec:beta_results}, a collected charge of \SI{20}{\femto \coulomb} ensures 30~ps or better time resolution for the sensors produced.  The operating voltage is taken to be the bias voltage at which the MPV of the collected charge reaches \SI{20}{\femto \coulomb}. In
Fig.~\ref{fig:beta_CV_measurements} (right) we show the CV transition voltage versus the operating voltage
for each of the sensors studied, where we observe a near linear relationship with few outliers. Therefore, it is possible to predict the operating voltage to within a few volts based on the measured CV transition voltage. This ability will be a crucial tool for designing the bias voltage distribution scheme and performing sensor quality control during the production of the ATLAS and CMS timing detectors.

For reliable operation of a large sensor, the pad with the smallest gain must reach the desired operating gain at a bias voltage less than the breakdown voltage of the pad with the largest gain. For the HPK type 3.1 batch, the difference between the operational bias voltage and the breakdown voltage is approximately 20~V as seen in Fig.~\ref{fig:beta_charge_vs_bias}. Furthermore, based on Fig.~\ref{fig:beta_CV_measurements}, all pads can reach the desired gain within a 20~V interval as long as the variation in the CV transition voltage is within approximately 1\% (the full range in the y-axis).

Probe station measurements have been previously reported over the scale of entire wafers that were part of this LGAD production~\cite{mferrero_CV}. 
The variation in the gain layer depletion voltage has been observed to be on the order of a few percent on distance scales across an entire wafer. However, in regions limited to the size of a single sensor for CMS and ATLAS (about \SI[product-units = power]{2 x 4}{\centi \meter}), the variation is limited to roughly 1\%. The results presented here shows that this magnitude of gain variation would allow all pads on a full sized sensor to be operated with better than \SI{30}{\pico \s} resolution even when constrained to a single bias voltage. We conclude that the gain layer uniformity achieved in HPK type 3.1 production batch would be sufficient to provide working full-sized sensors for CMS and ATLAS.

\section{Conclusion}
\label{sec:conclusion} 

We report comprehensive studies of HPK type 3.1 LGAD prototypes for the CMS and ATLAS timing detectors, including testbeam, beta source, and probe station measurements. Through careful design of each measurement campaign, we have been able to correlate the results from each measurement significantly expand the value of each technique. By comparing to testbeam measurements, we successfully validated the accuracy of the beta source measurements, which enables us to survey a much larger volume of sensors. Careful subsequent comparisons with probe station measurements allowed us to translate the impact of subtle variations in the gain layer doping concentration to quantitative variations in operating voltage. Ultimately this collection of measurements yields the possibility for detailed assessment of LGAD productions relying only on simple probe station measurements. 

The uniformity observed in the HPK 3.1 LGAD production is sufficient to produce working, full-sized sensors for CMS and ATLAS capable of yielding a time resolution of \SI{30}{\pico \s} before irradiation. This conclusion addresses one of the two critical questions facing the sensors for these timing detectors. The remaining question is to demonstrate that the LGAD sensors have sufficient radiation tolerance to survive until the end of the life of the HL-LHC, or a fluence up to $1.5 \times 10^{15} \mathrm{neq}/\si{\centi\m\squared}$ for the inner radius of CMS. The thoroughly characterized HPK 3.1 LGAD sensors documented in this paper provide an excellent sample for robust measurements of the radiation hardness with high statistics.

\section*{Acknowledgment}

We thank the Fermilab accelerator's team for very good beam conditions during our test beam time. We thank Mandy Kiburg, Evan Niner, Todd Nebel, Jim Wish, and all of the FTBF personnel for their support during the test beam experiments. We would like to thank Lorenzo Uplegger, Alan Prosser and Ryan Rivera for their critical role in establishing the FTBF tracker and its DAQ and trigger chain.  We are grateful for the technical support of the Fermilab SiDet department, especially Bert Gonzalez and Michelle Jonas for the rapid production of wire-bonded and packaged LGAD assemblies, and Abhishek Bakshi for the mechanical design of the experimental structures. 
This document was prepared using the resources of the Fermi National Accelerator
Laboratory (Fermilab), a U.S. Department of Energy, Office of Science, HEP User
Facility. Fermilab is managed by Fermi Research Alliance, LLC (FRA), acting
under Contract No. DE-AC02-07CH11359. Part of this work was performed within the
framework of the CERN RD50 collaboration.

This work was supported by the Fermilab LDRD 2017.027; by the United States
Department of Energy grant DE-FG02-04ER41286; by the California Institute of
Technology High Energy Physics under Contract DE-SC0011925; by the European
Union's Horizon 2020 Research and Innovation funding program, under Grant
Agreement no. 654168 (AIDA-2020) and Grant Agreement no. 669529 (ERC
UFSD669529); by the Italian Ministero degli Affari Esteri and INFN Gruppo Vl; and by the National Research Foundation of Korea (NRF) grant funded by the Korea government (MSIT) (Grants No. 2018R1A6A1A06024970, No. 2020R1A2C1012322 and Contract NRF-2008-00460).





\bibliographystyle{ieeetr} 
\bibliography{HPK3p1}

\begin{thebibliography}{10}

\bibitem{Apollinari:2120673}
G.~Apollinari, O.~Brüning, T.~Nakamoto, and L.~Rossi, ``{Chapter 1: High
  Luminosity Large Hadron Collider HL-LHC. High Luminosity Large Hadron
  Collider HL-LHC},'' {\em CERN Yellow Report}, pp.~1--19. 21 p, May 2017.

\bibitem{Schulte:2017qkc}
D.~Schulte, ``{FCC-hh Design Highlights},'' {\em ICFA Beam Dyn. Newslett.},
  vol.~72, pp.~99--109, 2017.

\bibitem{CMS:2667167}
{CMS Collaboration}, ``{A MIP Timing Detector for the CMS Phase-2 Upgrade},''
  Tech. Rep. CERN-LHCC-2019-003. CMS-TDR-020, CERN, Geneva, Mar 2019.

\bibitem{Cartiglia201783}
{N. Cartiglia, A. Staiano, V. Sola, \textit{et al.}}, ``Beam test results of a
  16 ps timing system based on ultra-fast silicon detectors,'' {\em Nucl.
  Instrum. Meth. A}, vol.~850, pp.~83 -- 88, 2017.

\bibitem{PELLEGRINI201412}
{G. Pellegrini, P. Fernandez-Martinez, M. Baselga, \textit{et al.}},
  ``{Technology developments and first measurements of Low Gain Avalanche
  Detectors (LGAD) for high energy physics applications},'' {\em Nucl. Instrum.
  Meth. A}, vol.~765, pp.~12 -- 16, 2014.

\bibitem{ApresyanLGAD}
{A.~Apresyan, S.~Xie, C.~Pena, \textit{et al.}}, ``Studies of uniformity of
  50~$\mu$m low-gain avalanche detectors at the fermilab test beam,'' {\em
  Nucl. Instrum. Meth. A}, vol.~895, pp.~158 -- 172, 2018.

\bibitem{JIN2020164611}
{Y. Jin, H. Ren, S. Christie, \textit{et al.}}, ``{Experimental Study of
  Acceptor Removal in UFSD},'' {\em Nucl. Instrum. Meth. A}, vol.~983,
  p.~164611, 2020.

\bibitem{Photek240}

\newblock
  \url{http://www.photek.com/pdf/datasheets/detectors/DS006_Photomultipliers.pdf}.

\bibitem{Keysight}

\newblock
  \url{https://www.keysight.com/en/pd-2233037-pn-MSOX92004A/infiniium-high-performance-oscilloscope-20-ghz}.

\bibitem{FTBF}

\newblock \url{http://ftbf.fnal.gov}.

\bibitem{KWAN2016162}
{S. Kwan, C.M. Lei, D. Menasce, \textit{et al.}}, ``{The pixel tracking
  telescope at the Fermilab Test Beam Facility},'' {\em Nucl. Instrum. Meth.
  A}, vol.~811, pp.~162--169, 2016.

\bibitem{4775101}
M.~{Turqueti}, R.~A. {Rivera}, A.~{Prosser}, J.~{Andresen}, and
  J.~{Chramowicz}, ``{CAPTAN: A hardware architecture for integrated data
  acquisition, control, and analysis for detector development},'' in {\em 2008
  IEEE Nuclear Science Symposium Conference Record}, pp.~3546--3552, 2008.

\bibitem{multiplexer}

\newblock \url{https://www.ni.com/en-us/support/model.pxi-2596.html}.

\bibitem{jarvis}

\newblock \url{https://github.com/CMS-MTD/JARVIS}.

\bibitem{BHARTHUAR2020164494}
{S. Bharthuar, J. Ott, K. Helariutta, \textit{et al.}}, ``{Study of
  interpad-gap of HPK 3.1 production LGADs with Transient Current Technique},''
  {\em Nucl. Instrum. Meth. A}, vol.~979, p.~164494, 2020.

\bibitem{PDG}
{M.~Tanabashi, \textit{et al.}}, ``Review of particle physics,'' {\em Phys.
  Rev. D}, vol.~98, Aug 2018.

\bibitem{Sola:2667026}
{V. Sola, R. Arcidiacono, N. Cartiglia, \textit{et al.}}, ``{Characterisation
  of 50~$\mu$m thick LGAD manufactured by FBK and HPK},'' {\em {14th
  “Trento” Workshop on Advanced Silicon Radiation Detectors FBK, Trento,
  25-27 February 2019}}, 2019.
\newblock \url{https://cds.cern.ch/record/2667026}.

\bibitem{mferrero_CV}
{M. Ferrero, \textit{et al.}}, ``{Recent studies and characterization on UFSD
  sensors}.'' 34th RD50 Workshop, Lancaster, UK, 2019.

\end{thebibliography}




\end{document}